\documentclass{aastex}

\usepackage{emulateapj5}
\usepackage{apjfonts}
\usepackage{amsmath}   
\usepackage{lscape}
\usepackage{portland}

\newcommand{\Vsys}{V$_{\mbox{\scriptsize sys}}$}
\newcommand{\Kp}{$K^\prime$}
\newcommand{\WRIi}{W$_{R,I}^{\mbox{\scriptsize i}}$}
\newcommand{\Vmax}{V$_{\mbox{\scriptsize max}}$}
\newcommand{\Vflat}{V$_{\mbox{\scriptsize flat}}$}
\newcommand{\HIsample}{{\em HI}--sample}
\newcommand{\SIsample}{{\em SI}--sample}
\newcommand{\RCsample}{{\em RC}--sample}
\newcommand{\RCRsample}{{\em RC/R}--sample}
\newcommand{\RCFsample}{{\em RC/F}--sample}
\newcommand{\RCDsample}{{\em RC/D}--sample}
\newcommand{\RCFDsample}{{\em RC/FD}--sample}
\newcommand{\RCFDnoNGCsample}{{\em RC/FD}$^{\mbox{\tiny $-$N3992}}$--sample}
\newcommand{\DEsample}{{\em DE}--sample}
\newcommand{\hf}{\hfill}
\newcommand{\nd}{\nodata}
\newcommand{\Xred}{$\chi^2_{\mbox{\scriptsize red}}$}
\newcommand{\sigdepth}{$\sigma_{\mbox{\scriptsize depth}}$}
\newcommand{\sigintr}{$\sigma_{\mbox{\scriptsize intr}}$}

\slugcomment{Accepted for publication in the Astrophysical Journal}

\shorttitle{The UMa Cluster of Galaxies. V ; Rotation Curves and the TF$-$Relations}
\shortauthors{M.A.W. Verheijen}

\begin{document}

\title{The Ursa Major Cluster of Galaxies. V ; HI Rotation Curve Shapes \\
       and the Tully-Fisher Relations.}
\author{Marc A.W. Verheijen}
\affil{Department of Astronomy, University of Wisconsin}
\affil{475 North Charter Street, Madison, WI 53706, U.S.A.}
\affil{verheyen@astro.wisc.edu}

\begin{abstract}

This paper investigates the statistical properties of the Tully-Fisher
relations for a volume limited complete sample of spiral galaxies in
the nearby Ursa Major cluster. The merits of B, R, I and \Kp\ surface
photometry and the availability of detailed kinematic information from
HI synthesis imaging have been exploited.  In addition to the
corrected HI global profile widths \WRIi, the available HI rotation
curves allow direct measurements of the observed maximum rotational
velocities \Vmax\ and the amplitudes \Vflat\ of the outer flat
parts. The dynamical state of the gas disks could also be determined
in detail from the radio observations.  The four luminosity and three
kinematic measures allowed the construction of twelve correlations for
various subsamples. For large galaxy samples, the
M$^{\mbox{\scriptsize b,i}}_{\mbox{\scriptsize R}}$$-$Log(\WRIi)
correlation in conjunction with strict selection criteria is preferred
for distance determinations with a 7\% accuracy.  Galaxies with
rotation curves that are still rising at the last measured point, lie
systematically on the low velocity side of the TF-relation. Galaxies
with a partly declining rotation curve (\Vmax~$>$~\Vflat) tend to lie
systematically on the high velocity side of the relation when using
\WRIi\ or \Vmax.  However, systematic offsets are eliminated when
\Vflat\ is used. Residuals of the M$^{\mbox{\scriptsize
b,i}}_{\mbox{\scriptsize B}}$$-$Log(2\Vflat) relation correlate
consistently with global galaxy properties along the Hubble sequence
like morphological type, color, surface brightness and gas mass
fraction. These correlations are absent for the near-infrared
M$^{\mbox{\scriptsize b,i}}_{\mbox{\scriptsize \Kp}}$$-$Log(2\Vflat)
residuals.  The tightest correlation (\Xred=1.1) is found for the
M$^{\mbox{\scriptsize b,i}}_{\mbox{\scriptsize \Kp}}$--Log(2\Vflat)
relation which has a slope of $-$11.3$\pm$0.5 and a total observed
scatter of 0.26 magnitudes with a most likely intrinsic scatter of
zero. The tightness of the near-infrared correlation is preserved when
converting it into a baryonic TF-relation which has a slope of $-$10.0
in case (${\cal M}$$_{\mbox{\scriptsize gas}}$/L$_{\mbox{\scriptsize
\Kp}}$)=1.6 while a zero intrinsic scatter remains most likely. Based
on the tightness of the near-infrared and baryonic correlations, it is
concluded that the Tully-Fisher relation reflects a fundamental
correlation between the mass of the dark matter halo, measured through
its induced maximum rotational velocity \Vflat, and the total baryonic
mass ${\cal M}$$_{\mbox{\scriptsize bar}}$ of a galaxy where ${\cal
M}$$_{\mbox{\scriptsize bar}}$$\propto$V$^4_{\mbox{\scriptsize
flat}}$.  Although the actual distribution of the baryonic matter
inside halos of similar mass can vary significantly, it does not
affect this relation.

\end{abstract}


\section{Introduction}

Numerous studies of the Tully-Fisher (TF) relations and their
applications have been conducted in the past. In most cases their aim
was to find observables that reduce the scatter and thus improve this
relation as a tool for measuring distances to spiral galaxies
(e.g. Peletier \& Willner 1993; Giovanelli et al. 1997).  However, the
statistical properties of the TF-relation are also used to constrain
numerical simulations of galaxy formation (e.g. Navarro \& Steinmetz
2000; Bullock et al. 2001), to probe galaxy evolution on cosmological
time scales (e.g. Vogt et al. 1997; Rix 1997), to determine
mass-to-light ratios of the stellar populations (e.g. Bell \& De Jong
2001), and to gain insights in the formation and structure of galaxies
(e.g. Courteau \& Rix 1999; Hinz, Rix, \& Bernstein 2001).

The rapid development of optical and near-infrared detector arrays has
led to greatly improved measurements of the luminosities of spiral
galaxies, ranging from the originally estimated photographic $B$
magnitudes \citep{tf77} to the present high quality near-infrared
surface photometry (e.g. Peletier \& Willner 1993; Tully et al. 1996;
Malhotra et al. 1996). The main advantages of near-infrared
luminosities are that stellar mass-to-light ratios are less affected
by population differences and that extinction corrections are minimal.
However, various studies yield conflicting results as far as the
scatter in the near-infrared relations is concerned. For example, for
a sample of Ursa Major spirals, \citet{pt88} found total scatters of
0.28$^m$ and 0.31$^m$ by using $I$ and $H_{-0.5}$ magnitudes
respectively.  \citet{bgr96}, using 23 spirals in Coma, also found
that the $H$-band TF-relation has no less scatter than the $I$-band
relation. On the other hand, \citet{msrl96} report scatters as low as
0.09, 0.13 and 0.20 magnitudes for the J, K and L passbands
respectively, using DIRBE maps for a sample of nearby galaxies
with Cepheid distances.

It is important to realize, however, that results from various
independent studies can not simply be compared at face value because
of different sample sizes and selection criteria, photometric
passbands, correction algorithms, fitting methods and statistical
techniques. For instance, the sample of \citet{msrl96} consists of
only seven galaxies and such small samples are known to yield
artificially low rms scatters when a least-squares fitting method is
applied (e.g. Freedman 1990). For cluster samples, the rms scatter
will be systematically larger due to the depth and velocity dipsersion
of the cluster.  As an extreme example, \citet{kkct88} reported a
scatter of 0.77$^m$ in the blue TF-relation for a sample of 127
galaxies in the Virgo cluster and it was shown by \citet{pt88} that
this large scatter was caused by contamination of fore- and background
galaxies which could not be identified and removed due to the
cluster's large velocity dispersion.  Two recent studies, one by
\citet{s00} and one by \citet{tp00}, used nearly identical samples of
nearby galaxies with independent measured Cepheid distances and both
studies included B, R and I band photometry. Nevertheless,
systematically different results were reported in the sense that
\citet{tp00} found shallower slopes, smaller scatters and brighter
zero points than \citet{s00} did, for each of those three
passbands. These differences can be largely attributed to different
extinction and other corrections.  Furthermore, from a statistical
point of view, it should be noted that the scatter is closely related
to the slope of the relation; a steeper slope gives a larger scatter
for the same uncertainties in the observables.

Most of the above-mentioned studies made use of the width \WRIi\ of
global HI profiles, corrected for instrumental resolution ($_R$),
internal turbulent motion ($_I$) and the inclination of the {\em
stellar} disk ($^{\mbox{\scriptsize i}}$). Although it is widely
recognized that the width of the HI profile is related to a galaxy's
rotation, relatively little attention has been given to what this HI
line width actually conveys. To some approximation, the rotational
velocity of a spiral galaxy might be derived from the properly
corrected width of its global HI profile. Indeed, this might be the
case if the rotation curve of the HI disk rises monotonically in the
inner regions and levels off to a constant velocity \Vflat\ in the
flat extended outer parts, yielding the characteristic double-horned
global profile. However, the width and detailed shape of a global HI
profile is determined by the shape of a galaxy's rotation curve, the
distribution of the HI gas in the disk and the possible presence of a
warp or non-circular motion.

From HI synthesis mapping of spirals it has become clear that there
are two basic deviations from this classical rotation curve shape.
First, many low surface brightness and dwarf galaxies only show the
rising part of the rotation curve; their HI disks do not extend far
enough to probe the regime of constant rotational velocity in the
halo. Consequently, their {\em observed} maximum rotational velocity
\Vmax\ provides merely a lower limit to the actual rotational velocity
induced by the potential of their dark matter halo. Second, the more
massive and compact galaxies often show a steep rise of the rotation
curve, which reaches a maximum \Vmax\ in the optical region followed
by a modest decline until \Vflat\ is reached in the outer flat parts
(e.g. Casertano \& Van Gorkom 1991). In this case \Vmax\ exceeds
\Vflat. The study presented in this paper aims at understanding the
statistical properties of the \WRIi-based TF-relation (tightness,
scatter \& slope) using knowledge of these detailed shapes of the HI
rotation curves and the characteristics of the full galaxy velocity
fields. Furthermore, it investigates whether such detailed kinematic
information can be used to improve the TF-relation as a distance
estimator and whether it is helpful in gaining a better understanding
of its physical basis.

The outline of this paper is as follows. Section 2 describes the
selected galaxy sample.  Section 3 provides details on the available
data. Section~4 explains the different shapes of the HI rotation
curves that can be identified.  In Section 5, various subsamples are
defined on the basis of the availability and characteristics of the
kinematic data.  Section 6 gives some special attention to the
corrections for internal extinction. Section 7 describes the
constructed TF-relations using the luminosities measured in the B, R,
I and \Kp\ passbands and the various kinematic measures \WRIi, \Vmax\
and \Vflat. The fitting method will be explained and the statistical
properties of the TF-relations will be examined given the shapes of
the rotation curves.  The residuals are examined in Section~8 in
search of a possible second parameter.  Section~9 discusses the
magnitude of any intrinsic scatter required to explain the total
observed scatter. In Section~10 the issue of the Low Surface
Brightness galaxies and their location on the TF-relation is briefly
discussed. Section~11 explores the properties of the baryonic
TF-relation. In Section~12 the results are discussed in relation to
other studies which consider the shapes of rotation curves. An attempt
is made to identify the fundamental parameters of a galaxy that form
the physical basis of the Tully-Fisher correlation. Finally,
Section~13 gives a summary of this paper.

\section{The sample}

The spiral rich and dynamically quiescent Ursa Major (UMa) cluster was
selected for the present study. An extensive description of this
cluster is given by \citet{tvphw96} (Paper I in this series). Paper I
also contains the photometric data and high quality blue and
near-infrared images of all the cluster members. The HI synthesis
imaging data of UMa cluster galaxies relevant for this Tully-Fisher
study are presented by \citet{vs01} (Paper IV).  Paper II in this
series \citep{tv97} investigates the optical and near-infrared central
disk surface brightness distributions. The high and low surface
brightness galaxies in the UMa cluster are identified in this paper.
For completeness, Paper III \citep{ttv01} deals with the faint end of
the R-band luminosity function in the UMa cluster.

Several characteristics of the Ursa Major sample have been updated
since Paper I was published. First of all, using newly available
Cepheid distances to local TF calibrator galaxies (e.g. Sakai et al.
2000), the distance to the Ursa Major cluster was revised from 15.5
Mpc in Paper I to 18.6 Mpc \citep{tp00} which is the distance adopted
in this paper. This new distance was also used in Papers III and
IV. Second, new corrections for internal extinction have been derived
by \citet{tphsvw98} based on a $\gamma$Log(b/a) scheme instead of the
\citet{tf85} (TFq hereafter) method which was used in Paper I. The
differences between these correction methods and their consequences
will be addressed later in this paper. Third, the CfA2 redshift
catalog has a deeper completion limit of m$_{\mbox{z}}$=15.5 instead
of the CfA1 limit of m$_{\mbox{z}}$=14.5 used to define the initial
complete sample in Paper I. Although the deeper CfA2 survey reveals no
new cluster members, the complete sample as defined in Paper I can now
be extended by 11 faint but known cluster members. The B, R, I and
K$^\prime$ photometry from Paper I is still nearly complete for this
extended complete sample but not all galaxies in the 14.5$-$15.5
magnitude range have HI synthesis data presented in Paper
IV. Therefore, the initially defined complete sample of galaxies
brighter than m$_{\mbox{z}}$=14.5 will be considered here with less
uncertainty about its completeness.  The updated distance of 18.6 Mpc
implies that all galaxies intrinsically brighter than
M$_{\mbox{\scriptsize B}}=-16.8$ are considered in this paper. This is
roughly twice the intrinsic luminosity of the Small Magelanic Cloud.

Sticking to the CfA1 completeness limit, a volume limited complete
sample of 62 galaxies can be constructed from the sample of 79 cluster
members identified in Paper I. However, not all of these galaxies are
suitable for a TF study due to their low inclinations. Therefore, a
lower inclination limit of 45$^\circ$, based on optical axes ratios and
an assumed intrinsic thickness of q$_0$=0.2, is imposed, reducing the
sample size from 62 to 49 galaxies. These 49 galaxies will be referred to as the
{\em complete} sample. Positional and redshift information on these
systems can be found in Table~1. \\
 \noindent{\it Column} (1) gives the NGC or UGC number. \\
 \noindent{\it Columns} (2,3) provide the equatorial coordinates in B1950. \\
 \noindent{\it Columns} (4,5) provide the Galactic coordinates. \\   
 \noindent{\it Columns} (6,7) contain the Super-Galactic longitude and
lattitude. \\
 \noindent{\it Column} (8) gives the systemic velocity
V$_{\mbox{\scriptsize sys}}$~$=$~V$_{\mbox{\scriptsize hel}}$ +
300sin(l)cos(b) where V$_{\mbox{\scriptsize hel}}$ is taken from Paper
IV unless noted otherwise. \\
 Note that the UMa cluster is located in the Super-Galactic plane at
about the same redshift as the core of the Virgo cluster. Figure~1
shows the distribution on the sky and in redshift space of the {\em
complete} sample. The cluster is diffuse and somewhat elongated along
the Super-Galactic plane without any central condensation or large scale
morphological segregation. There seems to be a trend of systemic
velocity with Super-Galactic longitude suggesting some substructure or
possible rotation of the galaxy ensemble.

\subsection{TF scatter due to depth effects}

Since all galaxies are nearly at the same distance, there is little
doubt about their relative luminosities, sizes and masses.  However,
the UMa cluster extends roughly 8$\times$15 degrees on the sky,
elongated along the Super-Galactic plane. Consequently, the depth of
the sample may be considerable. The depth of the cluster is estimated
by assuming that the distribution of all 79 identified cluster
galaxies along the line of sight is similar to the distribution along
Super-Galactic longitude. From this assumption it follows that the
estimated depth of the cluster contributes roughly \sigdepth=0.17$^m$
to the total observed scatter in the TF-relation.

This estimate of the depth is a conservative lower limit since the
Ursa Major cluster is located near the triple-value point of the
projected velocity field induced by the gravitational potential of the
Virgo cluster. Consequently, significant velocity crowding occurs in
the direction of the Ursa Major cluster and the cluster's velocity
range of 700$<$\Vsys$<$1210 km/s could cover as much as a factor 2.1
in distance between the near and the far side of the volume
(e.g. Figure~2 in Davis \& Peebles 1983). From this, a reasonable
upper limit to the depth of the cluster follows by stretching the
sample along this fully permitted line-of-sight, contributing a
maximum of \sigdepth=0.28$^m$ to the total observed scatter in the
TF-relation for a slope of $-$10. However, in the remainder of this
paper, \sigdepth=0.17$^m$ will be assumed. It should be noted that the
effect of velocity crowding could lead to some contamination by
galaxies, found in the cluster's limited velocity range, that are
located significantly more nearby or farther away than the main body
of the cluster.

\section{Available data}

Table~2 compiles the available relevant photometric and kinematic data
required to construct the various TF-relations as well as some global
parameters useful for searching for second parameters. \\
 \noindent{\it Column} (1) gives the NGC or UGC number.\\ 
 \noindent{\it Columns} (2-5) provide the total optical $B_J$, $R_{KC}$,
$I_{KC}$ and near-infrared \Kp\ magnitudes corrected for Galactic
and internal extinction as described in Paper IV. Total apparent
magnitudes were taken from Paper I. Uncertainties in the total apparent
magnitudes are estimated at 0.05 magnitudes in the $B$, $R$ and $I$
passbands and 0.08 magnitudes in the \Kp\ passband.  These
estimates are based on comparisons of the derived luminosities of
galaxies which were observed during different runs. \\
 \noindent{\it Column} (6) gives the adopted inclination and its
estimated uncertainty as motivated in Paper IV. \\
 \noindent{\it Columns} (7) contains the width of the global HI profile
corrected for instrumental resolution, random motions and inclination as
described in Paper IV. There are four S0 galaxies (N3990, N4111, N4143,
N4346) in which no HI could be detected. \\
 \noindent {\it Column} (8) contains information on the shape of the HI
rotation curves. See Section~4 for details. \\
 \noindent {\it Column} (9) provides the inclination corrected maximum
rotational velocities \Vmax\ measured from the HI rotation curves. For
galaxies with a rising `R-curve', \Vmax\ is the velocity of the last
measured point. For galaxies with a flat `F-curve', \Vmax\ is the
rotational velocity averaged over the outer flat part. For galaxies
with a declining `D-curve', \Vmax\ represents the peak velocity. \\
 \noindent {\it Column} (10) gives the inclination corrected amplitude
of the flat part of the rotation curve. For galaxies with a R-curve,
\Vflat\ could not be measured. For galaxies with a F-curve,
\Vflat~=~\Vmax. For galaxies with a D-curve, \Vflat\ is the rotational
velocity averaged over the outer flat part in which case
\Vflat~$<$~\Vmax. See Section~4 for a detailed explanation. \\
 \noindent {\it Column} (11) gives the morphological type of the galaxy. \\
 \noindent {\it Column} (12) contains the inclination corrected central
disk surface brightness in the \Kp\ band, taken from Paper II.
High Surface Brightness (HSB) galaxies are those galaxies with
$\mu^i_0(\mbox{\Kp})$~$<$~18.5 while Low Surface Brightness (LSB)
galaxies have $\mu^i_0(\mbox{\Kp})$~$>$~18.5. \\
 \noindent {\it Column} (13) gives the exponential disk scale lengths in
arcminutes, measured in the near-infrared. \\
 \noindent {\it Column} (14) provides the compactness parameter C$_{82}$
which is the ratio of the radii containing 80\% and 20\% of the total
light, averaged over the B, R and I bands. In some sense this is a
measure of the bulge-disk ratio. \\
 \noindent {\it Column} (15) gives the integrated HI flux density as
measured from the global HI profile. \\
 \noindent {\it Columns} (16-17) contain the IRAS 60 and 100 micron
fluxes taken from the IRAS Faint Source Catalog. Upper limits were
obtained using SCANPI, available at IPAC, and represent 5 times the
noise level in the median co-added local IRAS scans. Eleven galaxies are
unfortunately located in the infamous IRAS gap. \\
 \noindent {\it Column} (18) gives the 21-cm radio continuum flux
density taken from Paper IV. Flux densities and upper limits for
galaxies not listed in Paper IV were taken from the NVSS \citep{ccgyptb98}. \\
 \noindent {\it Column} (19) shows to which subsample each galaxy is
assigned. The various subsamples are explained in Section~5. \\
 \noindent {\it Column} (20) contains references to comments listed in
Table~3.

\section{HI Rotation Curve Shapes}

The available radio synthesis data presented in Paper IV provide HI
rotation curves which, for now, can be classified in three catagories
as illustrated in Figure~2.

First, there are rotation curves that keep on rising continuously
until the last measured point (left panel). In these cases, the {\em
observed} maximum rotational velocity is determined by the extent of
the HI disk and provides a lower limit to the actual maximum
rotational velocity induced by the gravitational potential. This type
of rotation curve is often observed in dwarf galaxies and in spirals
with an HI disk that is confined to the inner regions. The global HI
profiles of these galaxies are in general Gaussian or boxy and lack a
clear double peaked signature. These rising rotation curves will be
refered to as `R--curves' hereafter.

Second, there are the `classical' rotation curves which show a modest
rise in the inner regions and then gently bend over to reach a more or
less extended flat part in the outer regions of the galaxy (middle
panel). This behaviour is typical for Sb-Sd type spirals. In a
standard rotation curve decomposition, the amplitude of the flat part
corresponds to the maximum rotational velocity induced by the dark
matter halo (e.g.  Van Albada \& Sancisi 1986). The global profiles of
these galaxies usually show a clear double peaked profile and the
amplitude of the flat part can in general be well retrieved from the
width of the global HI profile. The selection criteria outlined by
\citet{bgrghhv94} are aimed at selecting galaxies with such a
classical rotation curve like N3917, based on their global properties.
These kinds of `classical' flat rotation curves will be referred to as
`F--curves'.

Third, there are the partly declining rotation curves which show a
steep rise in the inner regions and reach a peak rotational velocity
well within the stellar disk, often induced by a massive bulge
component (right panel). After this maximum they show a modest decline
within the observed stellar disk, typically beyond 2-3 disk scale
lengths. In most cases this decline halts and the rotation curve
extends further out with a more or less contant rotational
velocity. This behaviour of the rotation curves is often seen in
massive early-type spirals and in galaxies with a compact distribution
of their luminous matter (e.g. Casertano \& Van Gorkom, 1991).  In
these systems, a distinction can be made between the maximum
rotational velocity \Vmax\ of the peak in the rotation curve and the
amplitude of its more or less extended flat part \Vflat. These
rotation curves with \Vmax~$>$~\Vflat\ will be refered to as
`D--curves' hereafter.

Of course, there are many intermediate cases and cases where one side of
the galaxy does show a flat part while the other side is still rising.
There are also rotation curves that are kinematically lopsided in the
sense that one side is rising more steeply than the other side.

Figure~2 demonstrates that {\em the} rotational velocity of a spiral
galaxy is not very well-defined.  The horizontal arrows in Figure~2
indicate the rotational velocities as derived from the corrected
global HI profiles.  In the case of N3917, the amplitude of the flat
part can be retrieved well from the global profile. In the case of
N4389, the amplitude of the flat part can not be measured from the
width of the global profile simply because it is not sampled by the HI
disk which is confined to the inner regions of the galaxy. In the case
of N3992, the width of the global profile yields the maximum
rotational velocity \Vmax\ and not the amplitude \Vflat\ of the outer
flat part. This situation arises because the bulk of the HI gas is
located near the maximum rotational velocity.  In general, the
rotational velocities derived from the global profiles of galaxies
with declining rotation curves depend on where the bulk of the HI gas
is located along the rotation curve.

Note that \Vmax\ and \Vflat\ were not just measured from
position-velocity diagrams as depicted in Figure~2. They were measured
from the rotation curves determined by fitting tilted rings to the
2-dimensional velocity fields as explained in Paper~IV.

The examples in Figure~2 illustrate how the width of a global HI
profile is not only determined by the intrinsic maximum rotational
velocity of a galaxy but also by the actual shape of the rotation
curve and the distribution of the HI gas.  Even when the line width is
measured with great accuracy, it does not guarantee a precise
measurement of \Vmax\ or \Vflat. Note, however, that early-type
spirals are often excluded in studies of the TF-relation as a distance
tool since inclusion of these galaxies is known to increase the
scatter in the relation (e.g. Rubin et al. 1985; Pierce \& Tully
1988). From an HI perspective, an increased scatter for early type
systems might be related to their relative small HI content
(e.g. Figure~5 in Paper IV) which could give a patchy sampling of the
rotation curve along the kinematic major axis.

\section{Subsamples}

The availability, quality and characteristics of the HI velocity
fields and rotation curves warrant the construction of various
subsamples drawn from the {\em complete} sample. Furthermore, using
the TF-relation as an empirical distance estimator requires a
different approach to galaxy selection procedures compared to its use
for constraining galaxy formation scenarios. In the first case,
galaxies are, often subjectively, selected on the basis of the
regularity of their morphology and other global properties with the
sole purpose to minimize the scatter in the observed relation. In the
second case, searching for the underlying physical basis of the
TF-relation, it is important to recognize to what extent the HI
kinematics convey information about the gravitational potentials of
galaxies with a wide variety of global properties. With these
considerations in mind, the following subsamples were constructed.

\begin{center}
{\it a) The \HIsample}
\end{center}

There are 45 galaxies in the {\em complete} sample that have a
measured HI global profile. These 45 galaxies will be referred to as
the {\em HI}-sample. Most of these profiles are presented in Paper
IV. There is one single dish profile (N3870) while three global HI
profiles (N4026, N4117, U7129) were obtained with a new deep VLA HI
survey of parts of the Ursa Major Cluster. This new survey aims at
measuring the slope of the HI mass function at the faint end and its
results will be presented in a forthcoming paper in this series. The
\HIsample\ will be used to construct Tully-Fisher relations for the
largest possible sample, invoking no selection criteria while using
global HI profile widths only, thereby ignoring any detailed kinematic
information. One of the goals of this paper is to understand the
scatter in this \HIsample\ using information from the HI rotation
curves.

\begin{center}
{\it b) The \DEsample}
\end{center}

When using the TF-relation as an empirical {\em D}istance {\em
E}stimator one can not afford to obtain HI rotation curves of each
galaxy to investigate its kinematic state. Therefore, galaxies are
usually selected on the basis of their global optical morphology and
the shape of their global (single-dish) HI profiles. To determine how
the current Ursa Major sample would `perform' as a distance estimator,
galaxies were selected from the \HIsample\ according to the same
selection criteria as outlined by \citet{bgrghhv94}. In their
study of spirals around the Coma cluster they only selected
non-interacting galaxies of type Sb-Sd, with steep HI profile edges,
with smooth outer isophotes and without a prominent bar. These
criteria are somewhat subjective but they are based on global
morphological characteristics that do not require detailed kinematic
information on individual systems. Table~4 summarizes which galaxies
from the \HIsample\ are excluded by these criteria. There are 16
galaxies that survive the various selection criteria and those 16 will
be referred to as the \DEsample. Because evaluation of these criteria
is somewhat subjective, the reader is encouraged to examine the actual
data presented in Papers~I and IV.

\begin{center}
{\it c) The \SIsample}
\end{center}

There are 38 galaxies in the \HIsample\ for which useful {\em
S}ynthesis {\em I}maging data is available. These 38 galaxies will be
referred to as the {\em SI}-sample and their HI data are presented in
detail in Paper IV. From these synthesis imaging data, the kinematic
state of the galaxies could be determined and for most galaxies the
shape of their HI rotation curves could be measured. The galaxies
1135+48, N3870, N3896, N4026, N4117, U7129 and N4220 did not make it
from the \HIsample\ into the \SIsample\ because of confusion and
signal-to-noise issues.

\begin{center}
{\it d) The \RCsample}
\end{center}

Not all galaxies in the \SIsample\ are kinematically well-enough
behaved to allow the measurement of a reliable rotation curve which
assumes that the HI clouds are on stable orbits around the galaxy
center.  Although judging the kinematic state of a galaxy is somewhat
subjective, seven galaxies in the \SIsample\ were identified as being
too disturbed for determining a reliable rotation curve.

From the HI velocity fields and column density maps it is clear that
several galaxies in the \SIsample\ are involved in interactions
(N3769, N3893, U6818, U6973) evidenced by tidal HI tails and disturbed
velocity fields. N3718 is extremely warped while the Seyfert galaxy
N4051 is very lopsided. It seems very likely that the assumption of
stable closed orbits is violated in these systems. One galaxy (N4389)
only shows HI emission along its prominent bar in which case the
`rotation curve' merely reflects streaming motions and the bar pattern
speed. In these cases the HI kinematics is unlikely to be a useful
tracer of the gravitational potential of the galaxies and the rotation
curves of these galaxies cannot be determined with much
confidence. There are 31 galaxies in the \SIsample\ for which a useful
HI {\em R}otation {\em C}urve could be measured and these 31 galaxies
are referred to as the \RCsample.

Within the \RCsample, there are 9 galaxies with a R-curve that make up
the \RCRsample, there are 15 galaxies with a F-curve that make up the
\RCFsample\ and 7 galaxies with a D-curve that make up the \RCDsample.
Consequently, there are 22 galaxies in the \RCsample\ for which
\Vflat\ could be measured with some confidence.  These 22 galaxies
make up the combined \RCFDsample\ ({\em RC/F} + {\em RC/D}) and
include galaxies with Sa~--~Sm morphologies, barred galaxies and
galaxies with mild kinematically lopsided rotation curves in their
inner regions.

\section{Internal extinction corrections}

The correction for internal extinction is most uncertain and
significantly affects the slope, zero point and scatter of the
TF-relation in the optical passbands. Therefore, it deserves some
special attention. It is worth pointing out the differences between the
two most widely used correction methods to appreciate their influences
on the statistical properties of the TF-relations. One of the two main
correction schemes was proposed by TFq where

\begin{align*}
 \mbox{A}_\lambda^{\mbox{\scriptsize i}} = -2.5\;\mbox{Log}\;[\; & f \left(1+e^{-\tau_\lambda \mbox{sec(i)}}\right) \\
                      & + (1-2f)\left(\frac{1-e^{-\tau_\lambda \mbox{sec(i)}}}{\tau_\lambda \mbox{sec(i)}}\right) ]
\end{align*}

\noindent with $f$ the fraction of stars above and below a slab of
dust in which the fraction (1-2$f$) of the stars is homogeneously
mixed with the dust. The opacity of the face-on dust layer is given by
$\tau_\lambda$. For inclinations above 80 degrees,
$\mbox{A}_\lambda^{i>80} = \mbox{A}_\lambda^{i=80}$ motivated by the
idea that the stars behind the slab of dust become visible once the
disk is sufficiently inclined. This `TFq-model' also corrects for
internal extinction in face-on galaxies.  This is a more or less
physically motivated model and the values of $f$ and
$\tau_{\mbox{\scriptsize B}}$ were initially estimated by TFq at
$f=0.25$ and $\tau_{\mbox{\scriptsize B}}=0.55$. The Galactic
reddening law was used to adopt $\tau_\lambda$ in the other
passbands. This correction scheme, with updated values for $f$ and
$\tau_\lambda$, was used in Paper I to obtain extinction corrected
magnitudes.

The other main correction scheme (e.g. Giovanelli et al. 1994) is more
empirical in nature and takes the form

 $$\mbox{A}_\lambda^{\mbox{\scriptsize i$\rightarrow$ 0}} = -\gamma_\lambda\;\mbox{Log}(a/b) $$

where $\gamma_\lambda$ is wavelength dependent and ($a$/$b$) is the
observed optical axis ratio. Note that this model only corrects toward
a face-on situation; it does not correct for extinction in face-on
galaxies for which (a/b)=1. The values of $\gamma_\lambda$ are
determined by minimizing the observed scatter in the TF-relations, an
exercise performed by many investigators in the past. For instance,
\citet{tphsvw98} used the present UMa sample combined with a sample of
galaxies in Perseus-Pisces to find that $\gamma_\lambda$ also depends
on the absolute galaxy luminosity; brighter galaxies contain
relatively more obscuring dust than fainter galaxies. The absolute
magnitude can be expresed in terms of the distance independent line
width through the TF-relations themselves via an iterative
scheme. Eventually, $\gamma_\lambda$ can be expressed as

 $$\gamma_\lambda = \alpha_\lambda + \beta_\lambda(\;\mbox{Log W}^i_{R,I} - 2.5 \; ) $$

\noindent where the values of $\alpha_\lambda$ and $\beta_\lambda$ are
given in Section~2 of Paper IV and $\gamma_\lambda=0$ in case
W$^i_{R,I}<85$ km/s. Since $\gamma_\lambda$ depends on the line width,
a major consequence is that it affects the slope of the TF-relation
which becomes steeper when using this latter correction scheme
compared to the TFq-model.  Table~1 of Paper IV contains the applied
internal extinction corrections for each individual galaxy in each of
the four passbands based on the $-\gamma_\lambda\;\mbox{Log}(a/b)$
scheme.  Corrections can be as high as 1.40, 1.02, 0.82 and 0.20
magnitudes in the B, R, I and K$^\prime$ passbands respectively. Using
data from all four passbands, a `redder than Galactic' reddening law
was found and explained by \citet{tphsvw98}.

It will be obvious that for highly inclined systems, the uncertainties
in the internal extinctions at $B$, $R$, and $I$ become quite
substantial even without considering galaxy-to-galaxy variations in
(f,$\tau_\lambda$) or $\gamma_\lambda$. For instance, no distinction
is made between dusty high surface brightness galaxies and the
relatively dust-free low surface brightness galaxies of similar
W$^i_{R,I}$.

The difference between the two corrections schemes is illustrated in
Figure~3. The four lines in each panel correspond to the 4 different
passbands considered here. The upper most line represents the larger
B-band correction. In the left hand panel A$^{\mbox{\scriptsize
i}}_\lambda$ is plotted against inclination for the TFq-model. Note
the non-zero correction for face-on galaxies and the constant
correction for highly inclined systems. The middle panel shows
A$_\lambda^{\mbox{\scriptsize i$\rightarrow$0}}$ as a function of
($b$/$a$) for a corrected global HI line width of \WRIi$=400$
km/s. Note the zero correction for face-on systems.  The amplitudes of
the curves in the middle panel increase with increasing line
width. For a line width of 85 km/s or less, all the correction curves
become zero.  The right hand panel compares the two models over a
range of inclinations relevant for the current study. An intrinsic
thickness of q$_0$=0.1 was adopted to convert ($b$/$a$) into an
inclination angle.  Considering the B-band correction for edge-on
systems, differences between the two models can be as large as +0.5
magnitudes for massive systems and $-$1.4 magnitudes for dwarf
galaxies with \WRIi$<85$ km/s.  These considerations alone strongly
support the use of \Kp\ luminosities, thus minimizing the
uncertainties that come with the internal extinction corrections.

\section{TF-relations}

In this section the TF-relations will be discussed in the various
bandpasses using the different kinematic measures
\WRIi, \Vmax\ and \Vflat. 

First the applied methods of fitting and calculating the total observed
scatter will be described. Then the TF-relations for the {\em HI}-- and
{\em DE}--samples using the global HI profile widths will be discussed.
Next, the shapes of the rotations will be used to explain the larger
scatter and shallower slopes found in the {\em HI}--sample.

\subsection{Fitting method}

For each passband and kinematic measure a straight line of the form

 $$ \mbox{M} = a + b\;\mbox{Log(W)} $$

\noindent was fit to the data points where M is the corrected absolute
magnitude and Log(W) is either Log(\WRIi), Log(2\Vmax) or
Log(2\Vflat). Inverse least-squares fits were made by minimizing

 $$ \chi^2(a^\prime,b^\prime) 
  = \sum\limits_{i=1}^{N}
    \frac{\left[ \mbox{Log(W)}_i - (a^\prime + b^\prime \mbox{M}_i) \right]^2}
         {\sigma^2_{\mbox{\scriptsize Log(W)}_i} + b^{\prime^2}\sigma^2_{\mbox{\scriptsize M}_i}} $$

\noindent and thus taking errors in both directions into account. The
zero point $a$ and slope $b$ of the TF-relation can then be recovered
according to $a=-a^\prime/b^\prime$ and $b=1/b^\prime$. The error
$\sigma_{\mbox{\scriptsize M}}$ in the absolute magnitude does not only
account for measurement errors of the actual luminosity but also
accommodates contributions to the scatter from uncertainties due to the
depth of the sample and a possible intrinsic scatter:

 $$ \sigma^2_{\mbox{\scriptsize M}} 
  = \sigma^2_{\mbox{\scriptsize M$_{obs}^{b,i}$}} 
  + \sigma^2_{\mbox{\scriptsize depth}} 
  + \sigma^2_{\mbox{\scriptsize intr}} $$

\noindent As argued in Section 2, it can be expected that
\sigdepth$\approx$0.17 mag. However, in practice \sigdepth\ and
\sigintr\ can not be separated. Therefore, in calculating \Xred, it was
decided to keep \sigdepth=0 and \sigintr=0 initially. In Section 9 it
will be explored how much of this additional scatter can be allowed
for by \Xred.

After the inverse least-squares fits are made, a weighted total
observed rms scatter $\sigma_{\mbox{\scriptsize obs}}$ can be calculated
according to

$$ \sigma^2_{\mbox{\scriptsize obs}} 
 = \frac{ \sum\limits_{i=1}^ {N} w_i \left[\mbox{M}_i - \left( a + b\mbox{Log(W)$_i$}\right) \right]^2} 
        { \sum\limits_{i=1}^{N} w_i } $$

\noindent with the weights $w_i$ calculated according to

$$ w_i
 = 1 / (\sigma^2_{\mbox{\scriptsize M$_i$}}  
 + b^2\;\sigma^2_{\mbox{\scriptsize Log(W)$_i$}}) $$

This formalism assumes that the errors $\sigma_{\mbox{\scriptsize M}}$
and $\sigma_{\mbox{\scriptsize Log(W)}_i}$ are independent and of a
Gaussian nature. Unfortunately, this is not quite the case.

When calculating $\sigma_{\mbox{\scriptsize M$_i$}}$ and
$\sigma_{\mbox{\scriptsize Log(W)}_i}$, the following measurement
errors can be considered. First there are the photometric
uncertainties $\sigma_{\mbox{\scriptsize m$_{\mbox{T}}$}}$ in the
total apparent magnitude which are estimated at 0.05 magnitudes in the
optical passbands and 0.08 in the near-infrared band. These errors are
estimated by comparing total apparent magnitudes of galaxies observed
during several of the 14 observing runs (see Paper I). However,
uncertainties vary from galaxy to galaxy and the values quoted above
just reflect the rms scatter of the differences. Second, there are the
formal measurement errors $\sigma_{\mbox{\scriptsize W$_{20}$}}$ on
the observed HI line widths W$_{20}$. Because the signal-to-noise of
the global HI profiles varies significantly from galaxy to galaxy, the
formal measurement errors cover a wide range and, indeed, can be very
small for high signal-to-noise profiles with steep edges. The values
of $\sigma_{\mbox{\scriptsize W$_{20}$}}$ do not reflect differences
in line widths caused by different measurement techniques (see Paper
IV).  Third, the uncertainties on \Vmax\ and \Vflat\ as listed in
Table~2 are not classical Gaussian 1-sigma errors but rather reflect a
fiducial range of values allowed by the HI position-velocity
diagrams. Fourth, the inclinations and their uncertainties are
estimated using various different observations like optical axis
ratios, ellipticities of HI disks, tilted ring fits to velocity fields
and morphologies of dust lanes. Consequently, the uncertainties in the
inclination angles $\sigma_{\mbox{\scriptsize i}}$ are not Gaussian in
nature (e.g.  89$\pm$1 degrees).

Apart from the non-Gaussian observational errors, there are also
uncertainties associated with the models used to correct the raw data
for Galactic and internal extinction, instrumental velocity resolution
and turbulent motions of the HI gas. The parameters used in the models
to correct for internal extinction ($\alpha_\lambda$, $\beta_\lambda$)
and turbulent motions (W$_{c,20}$, W$_{t,20}$) have only a statistical
meaning (see Paper IV). Consequently, estimates of their uncertainties
are dictated by the concept of the model and lose their meaning in
any Gaussian error propagation exercise. One should also keep in mind
that the values of ($\alpha_\lambda$, $\beta_\lambda$) were determined
in an iterative way by minimizing the observed scatter in the
TF-relation in the first place. Finally, note that the values of
M$_{\mbox{\scriptsize T}}^{\mbox{\scriptsize b,i}}$ and Log(\WRIi) and
their errors are not entirely independent because
M$_{\mbox{\scriptsize T}}^{\mbox{\scriptsize b,i}}$ depends to some
extent on \WRIi\ through the correction for internal extinction; a
larger value of \WRIi\ implies a larger internal extinction
correction, depending on ($b$/$a$), and thus a brighter absolute
magnitude.

All these considerations make a rigorous statistical treatment of the
observed scatter in the TF-relation practically impossible and one
might argue against a 1/$\sigma^2$ weighting alltogether. Therefore,
in practice, inverse least-squares fits with equal weights (i.e. all
points have similar errors) were made by assuming that all galaxies
have equal relative uncertainties of 5\% in \WRIi, \Vmax\ or \Vflat\
and equal photometric uncertainties of 0.05 mag in
M$_{\mbox{\scriptsize T}}^{\mbox{\scriptsize b,i}}(B,R,I)$ and 0.08
mag in M$_{\mbox{\scriptsize T}}^{\mbox{\scriptsize b,i}}$(\Kp). These
estimated average uncertainties allow to calculate a fiducial value of
\Xred in order to evaluate the necessity of any intrinsic scatter to
explain the total observed scatter. Note that the adopted photometric
uncertainties in the optical absolute magnitudes are probably
underestimates given the applied corrections for internal extinction.

The total observed scatter was calculated according to 

$$ \sigma^2_{\mbox{\scriptsize obs}}
 = \frac
   { \sum\limits_{i=1}^{N} w_i \left[\mbox{M}_{\mbox{\scriptsize T}}^{\mbox{\scriptsize b,i}}(\mbox{obs})_i - \left( a + b\mbox{Log(W)$_i$}\right)\right]^2}
   { \sum\limits_{i=1}^{N} w_i }
$$

\noindent where $w_i$ was calculated as above and `$a$' and `$b$' are the
fitted zero point and slope of the TF-relation. In practice, $w_i$ is
identical for all galaxies.

Since the rms scatter may be strongly affected by outliers, it often
does not represent the scatter of the bulk of the data points.
Therefore, it is useful to calculate the more robust bi-weight scatter
$\sigma^{\mbox{\scriptsize bi}}$ \citep{bfg90} as well.  Values of
$\sigma^{\mbox{\scriptsize bi}}$ were determined after a normal
least-squares fit was made. For a pure Gaussian distribution of
residuals one would find $\sigma^{\mbox{\scriptsize
rms}}$$=$$\sigma^{\mbox{\scriptsize bi}}$.

Finally, it should be noted that a tighter TF-relation does not
necessarily imply a better distance tool. The tightness of the
relation is expressed by \Xred\ which is calculated by considering
errors in both directions. The usefulness of the TF-relation as a
distance tool, however, depends on the scatter in the direction of the
luminosity axis which is related to the steepness of the
slope. Consequently, a tighter correlation with a steeper slope may
still display a larger scatter along the luminosity axis.

The results of the inverse least-squares fits in the various passbands
for the various subsamples are listed in Table~5.

\subsection{TF-relations using global profiles.}

Figure~4 shows the TF-relations in each of the four passbands,
constructed with the corrected widths of the global profiles \WRIi.
First, the full \HIsample\ will be considered. In the upper panels,
all 45 galaxies in the \HIsample\ are plotted, regardless of their
morphologies, HI profile shapes and kinematic state. In the upper
\Kp-panel, U7129 is missing because it lacks a \Kp\ magnitude
measurement. Table~5 lists the results of the inverse least-squares
fits. The scatters are similar ($\sim 0.6^m$) in the four passbands
although the slope is steepening systematically from $-$6.8 to $-$8.0
going from the blue to the near-infrared. The total observed scatters
in the \HIsample\ are far larger than can be explained by the
estimated measurement uncertainties as suggested by \Xred$>>1$ while
there are no significant indications for curvature or other deviations
from a linear relation.  Note that the bi-weight scatter is slightly
smaller than the rms scatter, indicating that there are non-Gaussian
outliers in the distribution of residuals. One such outlier in the
optical passbands is N4117, a rather red dwarf lenticular galaxy (see
Paper I). It seems to be underluminous in the blue passband but it
systematically shifts toward the relation going further to the
(near-infra)red. The scatter in the \HIsample\ demonstrates that the
TF-relations are of limited use as distance estimators if no further
selection criteria are imposed.

The lower panels in Figure~4 show the TF relations for the 16 regular
galaxies in the restricted \DEsample\ who were selected on the basis
of their global morphological properties. Obviously, selecting
galaxies on the basis of their regular appearance greatly reduces the
scatter to 0.24$-$0.32 magnitudes, although \Xred\ indicates that the
scatters are still somewhat larger than can be expected from the
measurement uncertainties, even when taking the depth effect into
account. The rms and bi-weight scatters are similar for the \DEsample\
indicating that outliers are effectively eliminated. The systematic
steeping of the slope toward the near-infrared is maintained in the
\DEsample\ but the slopes have become steeper for each passband,
ranging from $-$8.2 to $-$10.6 going from the blue to the
near-infrared.  These steeper slopes are a result of the fact that the
selection criteria for the \DEsample\ eliminate galaxies in certain
areas of the TF-diagrams. As a consequence of the steeper slopes, the
zero points have changed significantly given the current formulation $
\mbox{M} = a + b\;\mbox{Log(W)} $ for the TF-relation (see Table~5).
The \DEsample\ also illustrates that a smaller scatter does not imply
a tighter relation.  Comparing the B- and \Kp-band relations, one
finds $\sigma_{\mbox{\scriptsize rms}}$=0.28 and \Xred=2.5 for the
B-band while $\sigma_{\mbox{\scriptsize rms}}$=0.32 and \Xred=1.9 for
the \Kp-band.

At this point, one can argue that the R-band TF-relation provides the
best distance estimator since it displays the smallest scatter. However,
it is not entirely obvious whether the {\em HI}-- or the \DEsample\
should be used to provide upper limits on the intrinsic scatter as a
constraint on galaxy formation scenarios, although the thightness of the
\Kp\ relation and the uncertainties related to the extinction
corrections in the optical seem to make the near-infrared relation more
relevant.

One of the main purposes of this paper is to investigate whether the
larger scatter and shallower slopes in the {\em HI}--sample can be
understood given the available detailed kinematic information in the
form of HI velocity fields and rotation curves. One might also want to
investigate whether the scatter in the {\em DE}--sample can be reduced
even further by using information from the HI rotation curves instead of
the width of the global HI profiles. These issues will be addressed in
the next subsection.

\subsection{TF-relations and the shapes of rotation curves}

Figure~5 shows the TF-relations in all four passbands, constructed for
the \RCsample\ of 31 galaxies which have measured rotation
curves.  The TF-relations in the upper panels were constructed using the
corrected widths of the global profiles \WRIi. The relations in the
middle panels were constructed using \Vmax\ from the rotation curves. 
The lower panels show the relations using the amplitude \Vflat\ of the
flat part of the rotation curve. The various symbols refer to the
shapes of the rotation curves: open triangles; galaxies with rising
R-curves (9 points), filled circles; galaxies with flat F-curves (15
points), and open circles; galaxies with declining D-curves (7 points).
Crosses indicate galaxies in the \SIsample\ that are excluded from the
\RCsample\ because they are involved in strong interactions or have
severe kinematic distortions which lead to an unreliable determination
of their rotation curves (7 points). The open triangles could not be
included in the lower panels simply because \Vflat\ can not be measured
for R-curves.

The results of the various inverse linear least-squares fits with
equal weights are collected in Table~5. As a reminder, weights are
based on a 0.05 mag error in the optical magnitudes, a 0.08 mag
error in the near-infrared magnitudes and a 5\% uncertainty in either
\WRIi, 2\Vmax\ or 2\Vflat . The values of \Xred\ quoted in Table~5 are
based on \sigdepth$^2$+\sigintr$^2$=0 in order to illustrate to what
extent the estimated observational errors can explain the total
observed scatter. Separate fits were made to the various {\em
RC}--subsamples, thereby ignoring the crosses. To allow a meaningful
comparison of the statistical properties of the correlation for the
three kinematic measures at a certain passband, one should consider
the same subsamples in all three of the rows.  Therefore, fits in the
upper and middle panels were also made considering only those galaxies
that appear in the lower panels for a certain subsample.

The solid lines in Figure~5 indicate the least-squares fits to the 15
filled symbols of the \RCFsample\ only. The dashed lines show the fits
to the entire \RCsample\ (31 galaxies in the upper and middle panels and
22 galaxies in the lower panels).  From Figure~5 and Table~5 the
following can be noted:

First of all, the 15 galaxies in the \RCFsample\ with a `classical'
rotation curve (F-curves with \Vmax~=~\Vflat, filled circles) define a
steeper and tighter relation (solid lines) compared to the entire
\RCsample\ (dashed lines). This happens to be the case in all passbands
using any of the three kinematic measures. However, when using \Vflat\
(lower panels) the differences between the solid and dashed lines are
insignificant. 

Second, galaxies with a rotation curve that is still rising at the last
measured point (R-curves with no \Vflat, open triangles) lie
systematically on the low velocity side of the relations defined by the
galaxies with F-curves (solid lines). These galaxies with R-curves are
mainly found among the fainter systems in the sample. If the HI disks of
those galaxies would have been more extended, they most likely would
have probed higher velocities further out into the halo and
consequently, those galaxies would have shifted to the right, i.e.
towards the relation. 

Third, galaxies with a partly declining rotation curve (D-curves with
\Vmax $>$\Vflat, open circles) in the upper and middle panels tend to
lie systematically on the high velocity side compared to galaxies with
\Vmax~=~\Vflat\ of the same luminosity.  If the lower amplitude
\Vflat\ of the flat part is used instead of the higher \Vmax\ values,
these galaxies shift to the left, i.e. toward the relation, and line
up with galaxies that have a `classical' F-curve as illustrated in the
lower panels.  Galaxies with declining rotation curves are mainly
found among the brightest systems in the sample and are in general of
earlier morphological type.

Fourth, the tightest relation with the smallest scatter is found for
the 15 galaxies in the {\em RC/F}--sample using \Kp\ magnitudes and
2\Vflat\ as the kinematic measure ($\sigma_{\mbox{\scriptsize
rms}}$=0.26 and \Xred=1.1). 

The M$^{\mbox{\scriptsize b,i}}_{\mbox{\scriptsize
\Kp}}$$-$Log(2\Vflat) relation for the 22 galaxies in the \RCFDsample\
gives a less tight correlation with a larger scatter.  However, this
can be entirely attributed to one single galaxy, NGC 3992, represented
by the most upper-right open circle in each panel of
Figure~5. Excluding N3992 from the \RCFDsample\ reduces the rms
scatter significantly from 0.32 to 0.26 magnitudes and \Xred\ from 1.8
to 1.1; the same statistical results are found for the smaller sample
of 15 galaxies in the \RCFsample. The reason why N3992 is not detected
as an outlier through the bi-weight scatter is the fact that it is at
the extreme end of the correlation and therefore it pulls the slope of
the least-squares fit towards it, thereby significantly reducing its
residual. Making an inverse fit to the M$^{\mbox{\scriptsize
b,i}}_{\mbox{\scriptsize \Kp}}$$-$Log(2\Vflat) correlation of the
\RCFDnoNGCsample\ and subsequently calculating the scatters {\em
including} N3992 yields $\sigma^{\mbox{\scriptsize rms}}$=0.35$^m$ and
$\sigma^{\mbox{\scriptsize bi}}$=0.31$^m$, a small but significant
difference. Removing the most deviating galaxy in the
\RCFDnoNGCsample\ (N3953, $\Delta$M=$-0.51^m$) does not significantly
reduce the scatter any further.

It is conceivable that N3992 is a background galaxy given its high
recession velocity of \Vsys$=1139$ km/s. Its companions U6923 and
U6969 also have high systemic velocities of 1151 and 1210 km/s
respectively, straddling the high velocity edge of the cluster's
700$-$1210 km/s velocity window. Furthermore, note that U6969, the
lower open triangle, also lies significantly below the
relation. Assuming that the N3992-group is 50\% farther away than the
UMa cluster as a whole, which is not inconceivable given the velocity
crowding in the UMa region, would put N3992 and U6969 back on the
TF-relation while U6923 moves slightly farther away from the relation,
making it more standing out as a galaxy with a R-curve (open
triangle).

Finally, removing N3992 from the \RCFDsample\ and considering the
remaining 21 galaxies in the {\em RC/FD$^{\mbox{\tiny
$-$N3992}}$}--sample, the correlation becomes robust, progressively
tighter and steeper and displays less scatter going from the blue to
the near-infrared (see Table~5). Furthermore, for the
\RCFDnoNGCsample, the correlation also tightens when using \Vflat\
instead of \Vmax\ from the rotation curve.

For the sceptical reader, Figure~6 provides a blow-up of the I-band
panel in the middle row of Figure~5. Each symbol is labeled with the
NGC or UGC numbers used in the various tables throughout this
paper. This allows for relating each individual data point to its
observational data presented in Papers I and IV.

\section{Searching for a 2nd parameter}

Eventhough \sigdepth=0, the least-squares fit results for the 21
galaxies in the \RCFDnoNGCsample\ show \Xred=1.1 in case of the
\Kp-band and no significant correlation of the residuals with other
parameters can be expected. However, in the B-band \Xred=4.0 which
leaves room for a possible second parameter to explain the excess of
scatter. It should be remarked that taking \sigdepth=0.17 mag into
account one finds \Xred$=$2.3 and \Xred$=$0.7 for the B-band and
\Kp-band relations respectively, indicating a possible slight
overestimation of the depth of the cluster or of the observational
errors in the M$^{\mbox{\scriptsize b,i}}_{\mbox{\scriptsize
\Kp}}$$-$Log(2\Vflat) relation. The values of \Xred\ quoted in the
remainder of this section are based on \sigdepth$^2$+\sigintr$^2$=0
unless explicitly mentioned.

Figure~7 shows the residuals in the M$^{\mbox{\scriptsize
b,i}}_{\mbox{\scriptsize B,\Kp}}$$-$Log(2\Vflat) TF-relations, as
displayed in the lower panels of Figure~5, plotted against various
global properties of the spirals. Only the $B$-band and \Kp-band
residuals are considered as extreme cases. As usual, solid symbols
indicate galaxies with F-curves and open symbols indicate galaxies
with D-curves. The crosses, indicating kinematically ill-behaved
systems in Figure~5, are omitted since they were not included in the
fits. The galaxy N3992 is also excluded from all the fits and is
indicated by an open symbol with a cross through it. Errorbars are
based on the estimated observational uncertainties and the slopes
fitted to the 21 galaxies in the \RCFDnoNGCsample. Note that there are
only 17 galaxies with a measured far-infrared flux.

The dashed lines in Figure~7 are direct least-squares fits to the
residuals, taking only the vertical errorbars into account. For each
panel, the number of points included in the fits as well as the
resulting values of the slopes and \Xred\ are listed in Table~6. The
galaxy U6399 is excluded from the fit in panel 3a because it clearly
seems to break the trend set by earlier type systems while no other
late type galaxies are available to confirm any turnover.

Figure~1 hinted at a possible correlation between \Vsys\ and SGL and
therefore, the TF residuals are plotted versus systemic velocity and
Super-Galactic Longitude in panels 1a,b and 2a,b of Figure~7. No
significant correlations are found with respect to the positions of the
galaxies in the cluster; the fitted slopes are not significantly
different from zero and the values of \Xred\ are not significantly
reduced. Nevertheless, the dashed line in panel 2b has a slope
consistent with the observation that galaxies at lower SGL have higher
\Vsys\ values (see Figure~1) and thus might be farther away in case they
are subjected to a Hubble flow. Consequently, they would be
underluminous (negative residuals) under the assumption that all
galaxies are at the same distance. However, \Xred\ is reduced from 1.1
to 0.8, indicating that the noise is being fit, while the slope deviates
only 2.3-sigma from zero. When invoking \sigdepth=0.17, \Xred\ reduces
further to 0.5 .

When inspecting Figure~7 and Table~6, it turns out that no significant
second parameter can be found for the \Kp\ residuals. However, for the
B-band residuals, several significant correlations with a second
parameter are evident in panels 3a, 5a, 7a and 8a. In those cases, the
fitted slopes deviate from zero by 5.0, 5.9, 6.2 and 4.3 sigma while
\Xred\ reduces from 4.0 to 2.4, 2.4, 2.2 and 3.3
respectively. However, most striking is the fact that these four
significant correlations are consistent with each other in the sense
that they reflect general trends along the Hubble sequence. Positive
residuals are found for late-type spirals which in general are gas
rich and bluer with a lower surface brightness. Effectively,
early-type spirals are too faint in their blue light or rotate too
fast compared to late type spirals. Because these correlations are
absent when using \Kp\ magnitudes, the B-band correlations can be
interpreted as the result of different stellar populations between
early and late type spirals of the same luminosity and not as the
results of differences in the rotational velocities \Vflat, induced by
their dark matter halos. If early-type spirals would reside in more
massive halos compared to their equiluminous late-type counterparts,
a similar correlation with morphological type would have been observed
for the \Kp\ residuals.

In light of the uncertain internal extinction corrections, it is
somewhat reassuring that no significant correlations are found between
the $B$-band and \Kp-band residuals with inclination.

Correlations of the residuals in the TF-relations with effective
surface brightness and compactness of the radial light distribution
were also found by \citet{w99}, using R-band magnitudes and H$\alpha$
long-slit rotation curves. He found that galaxies with a high
effective surface brightness or a highly compact light distribution
are too faint for their rotational velocity or rotate too fast for
their luminosity; they lie systematically below the TF-relation.
Interestingly, this is consistent with results from \citet{rbft85} who
found that early-type spirals lie systematically below the TF-relation
defined by later types (see also Section~12 in this paper).
\citet{cr99} analyzed the residuals from a mean relation between
parameterized velocities at 2.2 disk scale lengths and I- and r-band
luminosities. They presented `tentative evidence' that the residuals
correlate with disk scale lengths and that there is a `slight
tendency' for them to correlate with color in the sence that redder
disks with shorter scale lengths rotate faster than bluer and more
extended disks of the same I- or r-band luminosity. The correlations
found in these studies are consistent and in qualitative agreement
with the current Ursa Major results. However, due to the different
data characteristics and analysis techniques, a more quantitative
comparison is unfeasible.

\section{Intrinsic scatter}

While reading the previous sections, it may have become clear that the
scatter in the TF-relation can be dramatically decreased by applying
proper selection criteria and especially when detailed knowledge about
the kinematics of individual spirals is available.  However, the
intrinsic scatter in the TF-relation is an elusive concept and its
meaning depends on the context in which it is discussed.  It has been
shown that the statistical properties of the TF-relation are closely
related to the passband in which the luminosities are derived and to
the selection criteria by which means the galaxy sample is
constructed.  Therefore, extreme care should be taken if one tries to
relate the intrinsic scatter to a certain degree of non-circular
motions induced by non-spherical halos when the intrinsic scatter is
derived from a sample which is critically selected and optimized to
serve as a distance tool (e.g.  Franx \& De Zeeuw 1992).  On the
other hand it would be unfair to relate the intrisic scatter to the
degree of accuracy with which distances can be measured if the
intrinsic scatter is derived from a complete volume limited sample
without any further selection criteria applied.  For instance, the
critically selected \DEsample\ yields a much tighter correlation than
the volume limited \HIsample.

Given the estimated observational errors, the 95\% confidence
intervals for both $\sqrt{\mbox{\sigdepth}^2 + \mbox{\sigintr}^2}$ and
the intrinsic scatter \sigintr\ were calculated assuming
\sigdepth=0.17 as motivated in Section~2.1.  These intervals indicate
the minimum and maximum values of the (intrinsic) scatters that can be
accommodated within the total observed scatter. The results for the
various passbands, samples and kinematic measures are collected in
Table~7. The minimum scatter, the most likely scatter and the maximum
scatter are given for each entry.  Note that, in general, the
confidence intervals become broader with smaller numbers of galaxies
in a sample and that, in principle, \sigdepth\ and \sigintr\ can not
be separated.

\subsection{TF as a distance tool: the M$^{\mbox{\scriptsize b,i}}_{\mbox{\scriptsize B,R,I,\Kp}}$$-$Log(\WRIi) relations.}

When applying the Tully-Fisher relation as an empirical distance tool
on large samples of galaxies, to measure large scale cosmic velocity
fields for instance, it is impractical to obtain detailed kinematic
information in the form of HI rotation curves for tens of thousands of
galaxies. The WHISP project (Westerbork observations of neutral
Hydrogen in Irregular and SPiral galaxies) aims at gathering HI
synthesis data of about 3000 HI rich nearby systems. However, for all
practical purposes, single dish global HI profiles are
used. Consequently, the M$^{\mbox{\scriptsize b,i}}_{\mbox{\scriptsize
B,R,I,\Kp}}$$-$Log(\WRIi) relations should be considered for the
\DEsample\ if the TF-relation is to be evaluated as a distance tool.

From Table~5 it is clear that the the R-band yields the smallest
scatter of 0.24 magnitudes with a slope of $-$8.8$\pm$0.4 for the
\DEsample\ when using HI line widths. It follows from Table~7 that,
for an estimated \sigdepth=0.17, the intrinsic scatter is most likely
around 0.06 magnitudes in the R-band. This means that, ignoring any
measurement error, distances can be determined with an accuracy of
typically 3 percent if R-band luminosities and global HI profiles are
employed and the abovementioned corrections and selection criteria are
applied. It seems somewhat unexpected that the near-infrared relation
is a significantly worse distance estimater, providing a distance
accuracy of only 10 percent. Note that both the R-band and \Kp-band
relation have a similar tightness. The reason why the \Kp-relation
shows a larger scatter is its steeper slope.

\subsection{TF constraining galaxy formation scenarios: the M$^{\mbox{\scriptsize b,i}}_{\mbox{\scriptsize B,R,I,\Kp}}$$-$Log(2\Vflat) 
relations.}

When using the TF-relations to constrain galaxy formation scenarios,
it is undesirable to apply selection criteria; one would like to
examine a volume limited sample, yet pay attention to the meaning of
the observables. In numerical simulations of galaxy formation one
usually identifies the dark matter halos and measures their masses to
determine the rotational velocity of the embedded stellar
disk. However, it was demonstrated in Section~4 that observationally,
the width of the global HI profile is not always a good measure of
this rotational velocity induced by the mass of the dark matter halo,
i.e. the amplitude of the outer flat part of the rotation curve. One
should be extremely careful when selecting the apropriate samples and
observables.

For example, \citet{ns00} found that the results from their galaxy
formation simulations are in agreement with the statistical properties
of the \HIsample. They apropriately rejected the observed statistical
properties of the smaller \DEsample\ given the notion that those
galaxies are selected on the basis of their regular morphological
appearance. However, although the \HIsample\ is more complete, the
line widths \WRIi\ of many galaxies in the \HIsample\ do not relate to
\Vflat. For their purposes, \citet{ns00} should have considered the
statistical properties of the steeper and tighter
M$^{\mbox{\scriptsize b,i}}_{\mbox{\scriptsize \Kp}}$$-$Log(2\Vflat)
correlation for the \RCFDnoNGCsample; all galaxies should be
considered in the entire volume limited sample for which \Vflat\ could
actually be measured. The other galaxies in the Ursa Major volume are
strongly interacting, too poor in HI, too face-on or have rising
rotation curves for which \Vflat\ could not be measured. Note that
these galaxies were mainly rejected because of observational issues,
not because of their intrinsic properties, except for the interacting
systems for which \Vflat\ can not be related to the mass of the dark
matter halo without knowing the dynamics of the interaction.

Examining the results from the inverse least-squares fits to the
M$^{\mbox{\scriptsize b,i}}_{\mbox{\scriptsize
B,R,I,\Kp}}$$-$Log(2\Vflat) relations in Table~5 indicates, going from
the blue to the near-infrared, a progressively steeper relation with a
decreasing scatter and, above all, a decreasing \Xred. The
corresponding estimated intrinsic scatters are 0.31, 0.15, 0.12 and
0.00 magnitudes for the B, R, I and \Kp bands respectively, based on
\sigdepth=0.17. This implies that {\em no intrinsic scatter is
required to explain the observed scatter in the M$^{\mbox{\scriptsize
b,i}}_{\mbox{\scriptsize \Kp}}$$-$Log(2\Vflat) relation}.

\section{Low Surface Brightness galaxies and the TF-relation}

It was shown by \citet{zhbm95} that Low Surface Brightness (LSB) and
High Surface Brightness (HSB) galaxies of the same luminosity lie at
the same location in the M$^{\mbox{\scriptsize
b,i}}_{\mbox{\scriptsize B}}$$-$Log(\WRIi) TF-relation, at least
within the $\sim0.8^m$ scatter of their B-band relation.  Since LSB
galaxies tend to be bluer than their HSB counterparts of the same
total mass, this situation must change with passband. If Zwaan et al's
assertion is correct, one consequently would expect the LSB galaxies
to be underluminous with respect to the HSB galaxies in the \Kp -band
TF-relation.

Figure~8 shows the M$^{\mbox{\scriptsize b,i}}_{\mbox{\scriptsize
B,\Kp}}$$-$Log(2\Vflat) relations for the \RCFDnoNGCsample. The filled
circles indicate the HSB galaxies and the open circles correspond to
the LSB systems. The distinction between HSB and LSB is made at the
near-infrared central disk surface brightness of $\mu^i_0${\scriptsize
(\Kp)}=18.5$^m$ as motivated in paper II. The sample of 21 galaxies in
the \RCFDnoNGCsample\ contains 13 HSB and 8 LSB galaxies with average
B$-$\Kp\ colors of 3.28$\pm$0.36 (scatter) and 2.52$\pm$0.39 (scatter)
respectively. This difference, or offset, of 0.76 magnitudes should
become evident when comparing the B-band and the \Kp-band relations.

Although HSB and LSB galaxies of similar luminosity can be found in
the UMa sample, the surface brightness correlates rather strongly with
total luminosity. In the current sample, the overlap region between
HSB and LSB galaxies is limited (see Figure~8) which complicates a
direct comparison of the HSB and LSB subsamples. Nevertheless, panel
8a of Figure~7 does show a significant correlation between the
M$^{\mbox{\scriptsize b,i}}_{\mbox{\scriptsize B}}$$-$Log(2\Vflat)
residuals and the near-infrared central disk surface brightness. Such
a correlation does not exist for the M$^{\mbox{\scriptsize
b,i}}_{\mbox{\scriptsize \Kp}}$$-$Log(2\Vflat) residuals (panel
8b). Therefore, it should be concluded that the current Ursa Major
sample indicates {\em there does exist an offset between HSB and LSB
galaxies in the B-band TF-relation} in the sense that LSB galaxies are
overluminous compared to HSB galaxies with similar \Vflat, or they
rotate too slow compared to their B-band equiluminous HSB
counterparts. However, {\em no such offset could be detected in the
tighter near-infrared relation}. This implies that the offset between
HSB and LSB galaxies in the B-band TF-relation is caused by
differences in the mass-to-light ratios of their stellar
populations. The offset is luminosity based and not mass related; LSB
galaxies are indeed overluminous in the B-band. The fact that no
offset could be detected in the \Kp-band TF-relation indicates that
{\em \Vflat\ is unrelated to the gravitational potential of the
stellar component.}

It should be noted that the corrections for internal dust extinction
complicate the situation in the B-band somewhat since the applied
corrections depend on total luminosity (or line widths), regardless of
surface brightness; HSB and LSB galaxies of the same luminosity
received the same extinction correction (see Paper IV).
Observationally, however, LSB galaxies turn out to be virtually dust
free. The LSB galaxies in the Ursa Major sample are hardly detected by
IRAS and in general, no CO emission from LSB systems can be detected
\citep{bh98}. This is illustrated by the HSB/LSB galaxy pair
N4085/N3917 for which the relevant data is listed in Table~8. These
two galaxies are of similar morphological type and mass (\Vflat) but
of different surface brightness. Note that the observed color
m$^{\mbox{\scriptsize b}}_{\mbox{\scriptsize
B}}$$-$m$^{\mbox{\scriptsize b}}_{\mbox{\scriptsize \Kp}}$
(uncorrected for internal extinction) of the HSB galaxy N4085 is
somewhat redder (3.85) than that of the LSB galaxy (3.50). Most
importantly, the relative far-infrared luminosity of the HSB galaxy is
6-7 times higher than that of the LSB system, indicating that the HSB
galaxy is much dustier than the LSB system. Note also that the dust in
the LSB galaxy is colder than that in the HSB system which still
leaves the possibility that large amounts of undetected obscuring
colder dust may be present in N3917.

\section{The Baryonic TF-relation.}

The tightest correlation, with a slope of $b=-$11.3, is found between
the \Kp\ luminosity of a galaxy and \Vflat, the rotational velocity
induced by the dark matter halo. Since the \Kp\ light is thought to be
a good tracer of the stellar mass, one may wonder whether the near-infrared
TF-relation changes significantly if the {\em total} baryonic mass,
i.e. stars+gas, is considered instead of just the stellar mass.  This
issue has been previously explored by \citet{msbb00} who noted a
break in the TF-relation around 0.5\WRIi=90 km/s below which galaxies
are underluminous. They were able to restore a linear TF-relation by
taking the gas mass of the galaxies into account. This break in the
TF-relation is not evident in the current UMa sample because it lacks
galaxies with \Vflat$<$80 km/s. Nevertheless, the available UMa data
are accurate enough to expect significant differences between a
luminosity-based and a mass-based TF-relation for galaxies with
\Vflat$>$80 km/s.

To study this issue with the current sample, a baryonic luminosity, is
calculated for all the galaxies by converting their gas content
(HI+He) into stellar \Kp\ light according to L$_{\mbox{\scriptsize
\Kp,gas}}$=1.4${\cal M}$$_{\mbox{\scriptsize HI}}$/(${\cal
M}$$_{\mbox{\scriptsize gas}}$/L$_{\mbox{\scriptsize \Kp}}$). The
factor 1.4 accounts for the Helium mass fraction while the dust mass
is neglected. This \Kp\ luminosity of the gas is then added to the
extinction corrected stellar luminosity to yield a baryonic
luminosity; L$^{\mbox{\scriptsize b,i}}_{\mbox{\scriptsize
bar}}$=L$^{\mbox{\scriptsize b,i}}_{\mbox{\scriptsize
\Kp,stars}}$+L$_{\mbox{\scriptsize \Kp,gas}}$.

Because the fainter galaxies are relatively more gas rich (see Figure
5 of Paper IV), L$_{\mbox{\scriptsize \Kp,gas}}$ will add relatively
more to the stellar luminosity of the fainter galaxies and
consequently, the slope of the baryonic TF-relation will become
shallower than the $b=-11.3$ of the \Kp\ luminosity based
TF-relation. However, it is not guaranteed that the tightness of the
\Kp\ luminosity based TF-relation will be maintained.  Results from
the inverse least-squares fits to the baryonic L$_{\mbox{\scriptsize
\Kp,bar}}$$-$Log(2\Vflat) correlation are given in Table~9 for a range
of values for (${\cal M}$$_{\mbox{\scriptsize
gas}}$/L$_{\mbox{\scriptsize \Kp}}$).

As can be seen from Table~9, a slope of $b=-10.0$ is found when (${\cal
M}$$_{\mbox{\scriptsize gas}}$/L$_{\mbox{\scriptsize \Kp}}$)=1.6 is
assumed. Furthermore, it follows that the 95\% confidence interval for
$\sqrt{\mbox{\sigdepth}^2+\mbox{\sigintr}^2}$ is
(0.00$-$0.14$-$0.27). Consequently, in case \sigdepth=0.17, there is
most likely no intrinsic scatter. In conclusion, {\em the slope of
$b=-11.3$ that is found for the tightest luminosity based TF-relation,
the near-infrared M$^{\mbox{\scriptsize b,i}}_{\mbox{\scriptsize
\Kp}}$$-$Log(2\Vflat) correlation, is reduced to $b=-10.0$ if the
gaseous component is converted into starlight on the basis of (${\cal
M}$$_{\mbox{\scriptsize gas}}$/L$_{\mbox{\scriptsize \Kp}}$)=1.6 in
which case a zero intrinsic scatter remains most likely}.

\section{Discussion}

This study is the first one that considers a {\em complete volume
limited} cluster sample of galaxies for which HI velocity fields are
obtained as well as multi-band optical and near-infrared imaging
photometry.  As such, the UMa dataset is unique but nevertheless, the
current results should be compared to at least a few of many other
studies which relate the shape of the rotation curves to the
statistical properties of the TF-relation. These studies come in
several flavors, depending on how the rotation curves are measured,
either via HI synthesis observations or by means of optical
spectroscopy and either with long slit spectroscopy (or its
radio-equivalent) along the major axis or through observations of a
galaxy's full velocity field.  Each methodology has its limitations in
the extent of the interpretation of the data, However, it is most
important to keep in mind that optical measures of a galaxy's rotation
are always limited to the inner regions of a (star forming) stellar
disk, no matter which method is employed. Optical observations cannot
probe a galaxy's rotation in the outer regions beyond the detectable
stellar disk where the gravitational potential is dominated by the
dark matter halo. This realm can only be probed accurately by studying
the kinematics of the extended HI disk.

\citet[chapter 3]{b92} collected accurate HI rotation curves from the
literature for a sample of 21 field galaxies with HI velocity fields.
About half of these galaxies show a significantly declining part.
Using $B$-band magnitudes, he found scatters of 0.70 mag when using
global profile widths, 0.62 mag when using \Vmax\ and 0.55 mag when
using \Vflat with slopes of $-$7.4, $-$8.3 and $-$8.5 respectively.
The correction methods he applied are somewhat different from the ones
applied to the Ursa Major data used in this paper (see Paper IV for an
extensive comparison and discussion) and, unfortunately, it is not
clear which fitting method, direct or inverse, he applied.  Although
he found much larger scatters than are found for the current UMa
sample (he used a nearby field sample with significant distance
uncertainties), he found the same trends in the scatter and slope of
the relation.  In the current case, however, nearly identical slopes
are found when using W$^i_R$ and \Vmax, thanks to a more appropriate
value for the turbulent motion parameter W$_t$ (see Paper IV).
\citet{b92} also made short WSRT observations (nearly equivalent to
optical long-slit spectroscopy) of a sample of 48 nearby field
galaxies.  In this larger sample he did not identify galaxies with
rising, flat or declining rotation curves but he merely measured a
`representative' rotational velocity and subsequently found no
improvement in the TF-relation when using this velocity instead of the
global profile width; the observed scatters were 0.77 and 0.81 mag and
slopes of $-$6.6 and $-$6.3 respectively. His conclusion that a
reduction of the scatter can only be achieved with more information
about the inclination angle to be derived from full HI velocity fields
was an important motivation for the current UMa study.

Several years later, \citet{r96} supplemented the two samples of
Broeils who, in the meantime, had measured 7 more rotation curves
derived from full velocity fields.  Rhee himself analyzed short WSRT
observations of 60 more spirals.  With these larger samples of 28
velocity fields and 108 short observations, \citet{r96} reached
exactly the same conclusions as \citet{b92} did before: rotation
curves derived from `long-slit' HI observations are not helpful to
reduce the scatter due to a lack of information about the inclination
of a galaxy.  This inclination angle and its possible change with
radius can be retrieved from an HI velocity field, however, and that
is why they advocate the use of full galactic velocity fields for
deriving accurately the rotational velocity of a galaxy.

The investigations by Broeils and Rhee are the only ones so far that
exploit the advantages of HI synthesis observations in an attempt to
gain a better understanding of the statistical properties of the
TF-relation.  This is mainly because the reduction and interpretation
of HI synthesis data is very elaborate.  And indeed, studies using
optical observations occur more frequently.  The major advantage of
optical spectroscopy is the much higher spatial resolution of the data
compared to that of radio synthesis data.  The disadvantages of
optical spectroscopy, however, are the relatively low velocity
resolution, the limited radial coverage and the influence of obscuring
dust in the observed galaxies.  Nevertheless, many investigators used
H$\alpha$ long-slit spectroscopy to retrieve the rotation curve of a
galaxy (e.g.  Rubin et al. 1985; Courteau 1992; Mathewson et al. 1992;
Vogt et al. 1997; Raychaudhury et al. 1997; Willick 1999 and many
references therein).

\citet{rbft85} noted a morphological segregation in the $B$- and
$H$-band TF-relations constructed by using \Vmax\ from their optical
rotation curves.  They constructed separate TF-relations for Sa, Sb
and Sc type spirals selected from the nearby field and noted that,
although the slopes are similar for each morphological class, in the
$B$-band, earlier type spirals are offset toward higher rotational
velocities with respect to later type spirals.  This offset is less
but not zero when using $H$-band magnitudes.  Their rotation curves do
not reach beyond R$_{25}$ and a decline in the rotation curve beyond
this radius, as is seen in the case of N3992 (see Figure~2), would not
have been detected by them.  They did not have the possibility to
explore the shapes of the rotation curves in the outer regions.
Furthermore, they found rather large scatters of 0.7 mag in both
passbands after shifting the three separate relations, one for each
morphological bin, on top of each other.  These results of
\citet{rbft85} can be reconciled with the current findings by noting
that declining rotation curves are mainly found in early type spirals.
It was impossible for \citet{rbft85} to measure the rotation curve at
large enough distances from the center to reveal a significantly
declining part in the outer regions.  Apparently, in most cases, the
relevant kinematic information is found in the very outer parts of the
rotation curves, beyond R$_{25}$.

Recently, \citet{rbbg97} investigated the $I$-band TF-relation for a
sample of spirals in the Coma cluster using long-slit H$\alpha$
spectroscopy.  They found, for a sample of 25 carefully selected Sb-Sd
spirals, an unusually shallow slope of $-$5 and a remarkably low scatter
of only 0.14 mag, barely consistent with their observational
uncertainties.  They investigated whether the scatter could be reduced
even further by using the shape of the rotation curve as an extra
parameter.  They quantified the shape in terms of the steepness of the
rotation curve in the {\it inner} part and, not surprisingly, failed
to reduce the scatter any further.  Due to a lack of signal, they
were not able to measure the shape of the rotation curve in the outer
regions.

\citet{sbwm93} present an extensive study using Fabry-Perot techniques
to map the H$\alpha$ velocity fields of 75 spirals in 3 clusters.  By
fitting tilted-rings to the velocity fields, they derive rotation
curves from which they extract the circular velocity.  A comparison of
these circular velocities with the velocities derived from global HI
profiles shows that the H$\alpha$ velocities are often estimated too
low.  It can be noted from their published rotation curves that many
are still rising at their last measured points and for those cases, it
is likely that a more extended HI disk would show a significantly
faster rotation in the outer regions.  Unfortunately, they did not
check whether the deviations between the optically and HI derived
rotational velocities are related to the slope of the optical rotation
curves at the last measured points.  Anyway, using $I$-band
magnitudes, they constructed TF-relations for 7 galaxies in the Antlia
cluster and for 10 galaxies in the Hydra cluster.  From double
regression fits, they find scatters of 0.18 and 0.29 mag for the two
clusters with slopes of $-$8.8$\pm$0.8 and $-$9.8$\pm$0.8
respectively.

Reviewing these studies of the TF-relation which make use of available
rotation curves, illustrates that measuring rotation curves via optical
spectroscopy gives insufficient information about the shape of rotation
curves in the relevant outer regions of spirals. 

The two main applications of the TF-relation are measuring galaxy
distances and constraining galaxy formation scenarios. These
applications require different approaches for constructing the samples
(only Sc vs all types), collecting the data (global profiles vs
rotation curves), retrieving the appropriate measures from the data
(\Vmax\ vs \Vflat) and interpreting the results in relation to their
applications (scatter vs tightness). Without sufficient information
about the shapes of galaxy rotation curves, it becomes difficult to
reliably relate the observed scatter in the TF-relation to, for
instance, the results of numerical galaxy formation simulations. The
main problem is that {\em the} rotational velocity of a galaxy is
simply not well defined. The rotational velocity induced by the dark
matter halo (\Vflat) in the outer parts of a galaxy can often not be
retrieved from global profiles or optical and even HI rotation curves,
no matter how accurate the observable is determined. This is
especially the case for volume limited samples which contain a wide
range of galaxy morphologies and gas mass fractions.

Using the TF-relation as a distance estimator for large samples of
galaxies requires the use of global HI profiles which can be collected
efficiently with single dish telescopes. The current Ursa Major sample
confirms that invoking strict selection criteria significantly reduces
the scatter in the relation (\HIsample\ vs \DEsample).  Furthermore,
the smallest rms scatter is obtained when using the global profile
widths in conjunction with the R-band luminosities. The total observed
scatter of 0.24 magnitudes in the R-band includes contributions from
measurement uncertainties, the depth of the sample and any `intrinsic'
scatter. Removing the estimated contribution due to the depth of the
sample (0.17$^m$) and assuming that the measurement errors are typical
for other observational surveys (0.05$^m$ in m$_{\mbox{\scriptsize
R}}$ and 5\% in W$^{\mbox{\scriptsize i}}_{\mbox{\scriptsize R,I}}$)
suggests that galaxy distances can be measured with a typical accuracy
of about 7\%.

When using the TF-relation to constrain galaxy formation scenarios,
one should consider the tightness of the relation and not so much the
scatter along the magnitude axis. In this case one should also include
the widest possible range of galaxy morphologies. Furthermore, making
the observations compatible with the results from numerical
simulations requires the measurement of \Vflat. Considering all Ursa
Major galaxies for which \Vflat\ could be measured shows that the
tightest correlation is found when using near-infrared \Kp\
luminosities (N=21, $b$=$-$11.3, $\sigma_{\mbox{\scriptsize
obs}}$=0.26$^m$, \Xred=1.1, excluding N3992) with a most likely zero
intrinsic scatter and a maximum allowed intrinsic scatter of 0.21
magnitudes. The fact that the tightest correlation is found when using
\Kp\ luminosities suggests that possibly the baryonic content of the
dark matter halo is fundamentally related to the mass of the dark
matter halo instead of the luminosity of the stellar component. After
all, the \Kp\ luminosity is more closely related to the stellar mass
than the optical luminosities are. In the previous section it was
shown that the gaseous component can be converted into a \Kp\
luminosity which can be added to the stellar \Kp\ luminosity without
losing the tightness of the correlation. It does, however, result in
a shallower slope since the low luminosity spirals are overall more
gas rich. A slope of $-$10 results when assuming a mass-to-light ratio
of (${\cal M}$/L$_{\mbox{\scriptsize \Kp}}$)=1.6 for the stellar or
gaseous component.

All these considerations suggest that the TF-relation is fundamentally
a correlation between the mass of the dark matter halo, expressed by
\Vflat, and the total baryonic mass in that halo while dark matter
halos of the same mass are required to have an identical \Vflat. The
fact that HSB and LSB galaxies follow the same near-infrared
correlation and that the correlation based on \Vflat\ is tighter than
the correlation based on \Vmax, indicates that the actual distribution
of the baryonic mass inside the halo is irrelevant. Furthermore, for a
reasonable mass-to-light ratio of the gas, the correlation between
dark halo mass and baryonic mass has a slope of $-$10 with little room
for any intrinsic scatter. The slopes of the luminosity based
TF-relations and the scatter in the correlation in the optical
passbands depend on the star formation histories and the resulting
current stellar populations. The scatter in the B-band relation cannot
entirely be interpreted in terms of non-circular motions, tri-axial
halo potentials or other dynamical processes since the near-infrared
correlation is significantly tighter. It should be recalled that these
statements only apply to galaxies more massive than about twice the
Small Magelanic Cloud.

\section{Summary}

This study investigated the statistical properties of the Tully-Fisher
relations for a complete volume limited sample of galaxies in the
nearby Ursa Major cluster. Total luminosities in the optical $B$, $R$, $I$
and near-infrared \Kp\ passbands were used in conjunction with the
corrected global HI profile widths \WRIi\ and information on the
shapes of the HI rotation curves. The two most commonly used methods
to correct the observed luminosity for internal extinction by dust are
compared. Rising, flat and declining rotation curves could be
identified from which \Vmax\ and \Vflat\ could be measured and related
to \WRIi. The four luminosity and three kinematic measures allowed the
construction of twelve Tully-Fisher correlations. The statistical
properties of these twelve correlations were investigated for various
subsamples which were constructed on the basis of the kinematic states
of the galaxies and the practical applications of the
correlation. Inverse least-squares fits were made taking fiducial
errors in both directions into account.

When using the TF correlation as a distance estimator, the global
profile widths \WRIi\ and the R-band luminosities are the preferred
observables in conjunction with strict selection criteria on the basis
of overall morphology. The resulting correlation allows distance
determinations with an accuracy of about 7\%, taking typical
measurement errors into account. This is based on a subsample of 16
critically selected galaxies.

Dropping the selection criteria and considering the correlations for
larger subsamples using \WRIi\ or \Vmax\ showed that galaxies with
rising rotation curves lie on the low velocity side of the correlation
while galaxies with declining rotation curves tend to be offset to the
high velocity side. The subsample of galaxies with classical flat
rotation curves shows a steeper and tighter correlation. When using
\Vflat\ instead of \WRIi\ or \Vmax, the offset for galaxies with
declining rotation curves disappears. The tightest correlation is
found when using \Kp\ luminosities and the amplitude \Vflat\ of the
flat part of the rotation curve. The tightness of this
M$^{\mbox{\scriptsize b,i}}_{\mbox{\scriptsize \Kp}}$$-$Log(2\Vflat)
correlation with a slope of $-$11.3, suggests a most likely intrinsic
scatter of zero and an upper limit to the intrinsic scatter of 0.21
magnitudes at the 95\% confidence level.  The residuals in the less
tight blue M$^{\mbox{\scriptsize b,i}}_{\mbox{\scriptsize
B}}$$-$Log(2\Vflat) relation correlate with global properties along the
Hubble sequence. Positive residuals are found for bluer spirals of
lower surface brightness and later morphological type which have a
relatively larger gas mass fraction. Notably, in the blue passband,
low surface brightness galaxies do not follow the same correlation as
high surface brightness galaxies since the residuals in the
M$^{\mbox{\scriptsize b,i}}_{\mbox{\scriptsize B}}$$-$Log(2\Vflat)
relation do correlate with the central disk surface brightness.
However, these correlations of the residuals with properties along the
Hubble sequence disappear in the near-infrared.

A baryonic TF-relation using \Vflat\ is considered by converting the
gas reservoir of a galaxy into a \Kp\ luminosity, assuming a range of
values for (M$_{\mbox{\scriptsize gas}}$/L$_{\mbox{\scriptsize
\Kp}}$), and adding that to the stellar \Kp\ luminosity. A slope of
$-$10.0 is found in case (${\cal M}$$_{\mbox{\scriptsize
gas}}$/L$_{\mbox{\scriptsize \Kp}}$)=1.6 while the tightness of the
correlation is maintained with a most likely intrinsic scatter of zero
for the baryonic TF-relation.

These findings indicate that the TF-relation is fundamentally a
correlation between the maximum rotational velocity of the dark matter
halo (\Vflat) and the total baryonic mass inside that halo. The actual
distribution of the baryonic mass inside the dark matter halo is
irrelevant. The scatter in the optical passbands is mainly related to
varying star formation histories and internal dust extinctions.

\acknowledgements

I wish to thank Renzo Sancisi and Brent Tully for initiating the Ursa
Major cluster project and for their collaborative support. Much of
this work was carried out at the Kapteyn Institute in Groningen, the
Institute for Astronomy in Honolulu and the National Radio Astronomy
Observatory in Socorro. The Westerbork Synthesis Radio Telescope is
operated by the Netherlands Foundation for Research in Astronomy
(NFRA) with financial support from the Netherlands Organization for
Scientific Research (NWO). This research has been supported by the
Leids Kerkhoven-Bosscha Fonds, NATO Collaborative Research Grant
940271 and the National Science Foundation Grant AST-9970780.

%Table 1
\begin{deluxetable}{lrrrrrrr}
\tabletypesize{\scriptsize}
\tablewidth{88mm}
\setlength{\tabcolsep}{1.9mm}
\renewcommand{\arraystretch}{0.8}
\newcommand{\phf}{\phantom{$^3$}}

\tablecaption{Coordinates for all 49 members of the UMa Cluster
brighter than M$_B=-$16.8 and more inclined than 45$^\circ$.}

\tablehead{
\colhead{Name}             &
\colhead{R.A.}             &
\colhead{Dec.}             &
\colhead{l}                &
\colhead{b}                &
\colhead{SGL}              &
\colhead{SGB}              &
\colhead{V$_{\mbox{sys}}$} \\
\colhead{}               &
\colhead{\phn$^{h}$ \phn$^{m}$ \phn$^{s}$\phd} &
\colhead{\phn$^\circ$ \phn$^\prime$ \phn$^{\prime\prime}$\phd} &
\colhead{deg}            &
\colhead{deg}            &
\colhead{deg}            &
\colhead{deg}            &
\colhead{km/s}           \\
\colhead{(1)}            &
\colhead{(2)}            &
\colhead{(3)}            &
\colhead{(4)}            &
\colhead{(5)}            &
\colhead{(6)}            &
\colhead{(7)}            &
\colhead{(8)}            }

\startdata
U6399   & 11 20 35.9 & 51 10 09 & 152.08 & 60.96 & 62.02 & $-$1.53 &  860\phf \\
U6446   & 11 23 52.9 & 54 01 21 & 147.56 & 59.14 & 59.72 &    0.22 &  727\phf \\
N3718   & 11 29 49.9 & 53 20 39 & 147.01 & 60.22 & 60.71 &    0.72 & 1074\phf \\
N3726   & 11 30 38.7 & 47 18 20 & 155.38 & 64.88 & 66.21 & $-$1.79 &  919\phf \\
N3729   & 11 31 04.9 & 53 24 08 & 146.64 & 60.28 & 60.74 &    0.91 & 1142\phf \\
N3769   & 11 35 02.8 & 48 10 10 & 152.72 & 64.75 & 65.74 & $-$0.75 &  796\phf \\
1135+48 & 11 35 09.2 & 48 09 31 & 152.71 & 64.77 & 65.75 & $-$0.74 &  798\phf \\
U6667   & 11 39 45.3 & 51 52 32 & 146.27 & 62.29 & 62.67 &    1.47 & 1051\phf \\
N3870   & 11 43 17.5 & 50 28 40 & 147.02 & 63.75 & 64.17 &    1.42 &  822\phf \\
N3877   & 11 43 29.3 & 47 46 21 & 150.72 & 65.96 & 66.68 &    0.38 &  955\phf \\
U6773   & 11 45 22.1 & 50 05 12 & 146.89 & 64.27 & 64.67 &    1.57 &  995\phf \\
N3893   & 11 46 00.2 & 48 59 20 & 148.15 & 65.23 & 65.73 &    1.24 & 1034\phf \\
N3896   & 11 46 18.6 & 48 57 10 & 148.10 & 65.29 & 65.78 &    1.27 &  913\phf \\
N3917   & 11 48 07.7 & 52 06 09 & 143.65 & 62.79 & 62.97 &    2.74 & 1046\phf \\
U6818   & 11 48 10.1 & 46 05 09 & 151.76 & 67.78 & 68.54 &    0.47 &  862\phf \\
N3949   & 11 51 05.5 & 48 08 14 & 147.63 & 66.40 & 66.83 &    1.70 &  865\phf \\
N3953   & 11 51 12.4 & 52 36 18 & 142.21 & 62.59 & 62.68 &    3.36 & 1137\phf \\
U6894   & 11 52 47.3 & 54 56 08 & 139.52 & 60.63 & 60.59 &    4.43 &  944\phf \\
N3972   & 11 53 09.0 & 55 35 56 & 138.85 & 60.06 & 59.98 &    4.72 &  951\phf \\
U6917   & 11 53 53.1 & 50 42 27 & 143.46 & 64.45 & 64.61 &    3.06 &  988\phf \\
N3985   & 11 54 06.4 & 48 36 48 & 145.94 & 66.27 & 66.56 &    2.34 & 1016\phf \\
U6923   & 11 54 14.4 & 53 26 19 & 140.51 & 62.06 & 62.07 &    4.09 & 1151\phf \\
N3992   & 11 55 00.9 & 53 39 11 & 140.09 & 61.92 & 61.91 &    4.27 & 1139\phf \\
N3990   & 11 55 00.3 & 55 44 13 & 138.25 & 60.04 & 59.95 &    5.01 &  696$^1$ \\
N4013   & 11 55 56.8 & 44 13 31 & 151.86 & 70.09 & 70.77 &    1.06 &  880\phf \\
N4010   & 11 56 02.0 & 47 32 16 & 146.68 & 67.36 & 67.69 &    2.26 &  966\phf \\
U6969   & 11 56 12.9 & 53 42 11 & 139.70 & 61.96 & 61.92 &    4.46 & 1210\phf \\
U6973   & 11 56 17.8 & 43 00 03 & 153.97 & 71.10 & 71.94 &    0.68 &  743\phf \\
U6983   & 11 56 34.9 & 52 59 08 & 140.27 & 62.62 & 62.61 &    4.26 & 1170\phf \\
N4026   & 11 56 50.7 & 51 14 24 & 141.94 & 64.20 & 64.27 &    3.68 & 1023$^1$ \\
N4051   & 12 00 36.4 & 44 48 36 & 148.88 & 70.08 & 70.51 &    2.04 &  753\phf \\
N4085   & 12 02 50.4 & 50 37 54 & 140.59 & 65.17 & 65.16 &    4.37 &  826\phf \\
N4088   & 12 03 02.0 & 50 49 03 & 140.33 & 65.01 & 65.00 &    4.46 &  838\phf \\
U7089   & 12 03 25.4 & 43 25 18 & 149.90 & 71.52 & 71.99 &    2.05 &  818\phf \\
N4100   & 12 03 36.4 & 49 51 41 & 141.11 & 65.92 & 65.93 &    4.23 & 1151\phf \\
U7094   & 12 03 38.5 & 43 14 05 & 150.14 & 71.70 & 72.19 &    2.02 &  827\phf \\
N4102   & 12 03 51.3 & 52 59 22 & 138.08 & 63.07 & 62.99 &    5.29 &  937\phf \\
N4111   & 12 04 31.0 & 43 20 40 & 149.53 & 71.69 & 72.13 &    2.20 &  815$^1$ \\
N4117   & 12 05 14.2 & 43 24 17 & 149.07 & 71.72 & 72.13 &    2.35 &  998\phf \\
U7129   & 12 06 23.6 & 42 01 08 & 151.00 & 72.99 & 73.47 &    2.11 &  977$^2$ \\
N4138   & 12 06 58.6 & 43 57 49 & 147.29 & 71.40 & 71.70 &    2.83 &  946\phf \\
N4143   & 12 07 04.6 & 42 48 44 & 149.18 & 72.40 & 72.79 &    2.47 &  985$^3$ \\
N4157   & 12 08 34.2 & 50 45 47 & 138.47 & 65.41 & 65.33 &    5.27 &  857\phf \\
N4183   & 12 10 46.5 & 43 58 33 & 145.39 & 71.73 & 71.90 &    3.48 &  984\phf \\
N4218   & 12 13 17.4 & 48 24 36 & 138.88 & 67.88 & 67.81 &    5.27 &  804\phf \\
N4217   & 12 13 21.6 & 47 22 11 & 139.90 & 68.85 & 68.82 &    4.96 & 1097\phf \\
N4220   & 12 13 42.8 & 48 09 41 & 138.94 & 68.13 & 68.07 &    5.26 &  988\phf \\
N4346   & 12 21 01.2 & 47 16 15 & 136.57 & 69.39 & 69.30 &    6.17 &  783$^1$ \\
N4389   & 12 23 08.8 & 45 57 41 & 136.73 & 70.74 & 70.65 &    6.16 &  786\phf \\
\enddata
\tablecomments{
$^1$: redshift from RC3.
$^2$: redshift from \citet{fkghpbmte99}
$^3$: redshift from \citet{dggp95}}
\end{deluxetable}

%Table~2
\begin{deluxetable}{lcccclcccclcccccclll}
\tabletypesize{\scriptsize}
\rotate
\tablewidth{220mm}
\setlength{\tabcolsep}{1.4mm}
\renewcommand{\arraystretch}{1.05}

\tablecaption{Various properties of the {\it complete sample}.}
\hf = \hfill
\newcommand{\tlt}{{\tiny$<$}}
\newcommand{\pht}{\hspace{3mm}}

\tablehead{
\colhead{Name}                                &
\colhead{M$^{\mbox{\scriptsize b,i}}_{\mbox{\scriptsize B}}$}          &
\colhead{M$^{\mbox{\scriptsize b,i}}_{\mbox{\scriptsize R}}$}          &
\colhead{M$^{\mbox{\scriptsize b,i}}_{\mbox{\scriptsize I}}$}          &
\colhead{M$^{\mbox{\scriptsize b,i}}_{\mbox{\scriptsize K$^\prime$}}$} &
\colhead{Incl.}                               &
\colhead{W$^i_{R,I}$}                         &
\colhead{Shape}                               &
\colhead{V$_{\mbox{\scriptsize max}}$}        &
\colhead{V$_{\mbox{\scriptsize flat}}$}       & 
\colhead{Type}                                &
\colhead{$\mu^i_0${\scriptsize (K$^\prime$)}} &
\colhead{h$_{\mbox{\scriptsize K$^\prime$}}$} &
\colhead{C$_{\mbox{\scriptsize 82}}$}         &
\colhead{$\int$Sdv}                           &
\colhead{F$_{\mbox{\scriptsize 60$\mu$m}}$}   &
\colhead{F$_{\mbox{\scriptsize 100$\mu$m}}$}  &
\colhead{F$_{\mbox{\scriptsize 1.4GHz}}$}     &
\colhead{Subsamples}                          &
\colhead{Comm.}                               \\
\colhead{}                                    &
\colhead{mag}                                 &
\colhead{mag}                                 &
\colhead{mag}                                 &
\colhead{mag}                                 &
\colhead{deg}                                 &
\colhead{km/s}                                &
\colhead{}                                    &
\colhead{km/s}                                &
\colhead{km/s}                                & 
\colhead{}                                    &
\colhead{mag/"}                               &
\colhead{($^\prime$)}                         &
\colhead{}                                    &
\colhead{Jy km/s}                             &
\colhead{Jy}                                  &
\colhead{Jy}                                  &
\colhead{mJy}                                 &
\colhead{HI SI RC DE}                         &
\colhead{}                                    \\
\colhead{(1)}  &
\colhead{(2)}  &
\colhead{(3)}  &
\colhead{(4)}  &
\colhead{(5)}  &
\colhead{(6)}  &
\colhead{(7)}  &
\colhead{(8)}  &
\colhead{(9)}  &
\colhead{(10)} &
\colhead{(11)} &
\colhead{(12)} &
\colhead{(13)} &
\colhead{(14)} &
\colhead{(15)} &
\colhead{(16)} &
\colhead{(17)} &
\colhead{(18)} &
\colhead{(19)} &
\colhead{(20)} }

\startdata
U6399   & $-$17.56 & $-$18.44 & $-$18.77 & $-$20.33 & 75$\pm$2 &    172$\pm$\phn2 & R/F   \hf\null & \phn88$\pm$\phn5 & \phn88$\pm$\phn5 & Sm   & 20.12 & 0.44 & 3.3 &    \phn10.5$\pm$0.3 & \tlt0.17 \hf\null & \tlt0.43 \hf\null & \phd\tlt2.5         & \null\hf $\times$ \hf $\times$ \hf   F   \hf  \pht    \hf\null &        \\
U6446   & $-$18.08 & $-$18.72 & $-$18.90 & $-$19.88 & 51$\pm$3 &    174$\pm$\phn8 & F     \hf    L & \phn82$\pm$\phn4 & \phn82$\pm$\phn4 & Sd   & 19.84 & 0.33 & 4.1 &    \phn40.6$\pm$0.5 & \tlt0.23 \hf\null & \tlt0.52 \hf\null & \phd\tlt7.2         & \null\hf $\times$ \hf $\times$ \hf   F   \hf $\times$ \hf\null &        \\
N3718   & $-$20.90 & $-$21.99 & $-$22.54 & $-$24.00 & 69$\pm$3 &    476$\pm$10    & F     \hf\null &    232$\pm$11    &    232$\pm$11    & Sa   & 18.45 & 0.94 & 4.6 &       140.9$\pm$0.9 & \phn0.70$\pm$0.07 & \phn2.29$\pm$0.27 &    \phn11.4$\pm$0.4 & \null\hf $\times$ \hf $\times$ \hf \pht  \hf  \pht    \hf\null & 12     \\
N3726   & $-$20.76 & $-$21.67 & $-$22.07 & $-$23.45 & 53$\pm$2 &    330$\pm$\phn9 & F/(D) \hf\null &    162$\pm$\phn9 &    162$\pm$\phn9 & SBc  & 17.72 & 0.63 & 3.1 &    \phn89.8$\pm$0.8 & \phn3.50$\pm$0.21 &    16.95$\pm$0.85 &    \phn49.7$\pm$5.0 & \null\hf $\times$ \hf $\times$ \hf   F   \hf  \pht    \hf\null & 6      \\
N3729   & $-$19.34 & $-$20.62 & $-$21.22 & $-$22.78 & 49$\pm$3 &    295$\pm$14    & F     \hf\null &    151$\pm$11    &    151$\pm$11    & SBab & 16.86 & 0.31 & 3.5 & \phn\phn5.5$\pm$0.3 & \phn2.68$\pm$0.16 & \phn7.38$\pm$0.44 &    \phn18.0$\pm$0.9 & \null\hf $\times$ \hf $\times$ \hf   F   \hf  \pht    \hf\null &        \\
N3769   & $-$19.32 & $-$20.35 & $-$20.80 & $-$22.35 & 70$\pm$2 &    257$\pm$\phn8 & F/(D) \hf\null &    122$\pm$\phn8 &    122$\pm$\phn8 & SBb  & 17.62 & 0.33 & 3.6 &    \phn62.3$\pm$0.6 & \phn2.63$\pm$0.13 & \phn6.78$\pm$0.41 &    \phn12.1$\pm$2.9 & \null\hf $\times$ \hf $\times$ \hf \pht  \hf  \pht    \hf\null & 7      \\
1135+48 & $-$16.67 & $-$17.51 & $-$17.88 & $-$19.40 & 73$\pm$3 &    111$\pm$\phn2 & confused       &    \nd           &    \nd           & Sm   & 20.11 & 0.23 & 3.6 & \phn\phn6.6$\pm$0.1 & confused          & confused          & \phd\tlt1.6         & \null\hf $\times$ \hf  \pht    \hf \pht  \hf  \pht    \hf\null & 7      \\
U6667   & $-$17.83 & $-$18.87 & $-$19.18 & $-$20.65 & 89$\pm$1 &    167$\pm$\phn2 & R/F   \hf    L & \phn86$\pm$\phn3 & \phn86$\pm$\phn3 & Scd  & 20.44 & 0.54 & 3.3 &    \phn11.0$\pm$0.4 & \tlt0.09 \hf\null & \tlt0.76 \hf\null & \phd\tlt2.7         & \null\hf $\times$ \hf $\times$ \hf   F   \hf $\times$ \hf\null &        \\
N3870   & $-$17.83 & $-$18.74 & $-$19.26 & $-$20.64 & 48$\pm$3 &    127$\pm$23    & \nd            &    \nd           &    \nd           & S0a  & 17.44 & 0.12 & 7.3 & \phn\phn5.6$\pm$1.6 & \phn1.09$\pm$0.07 & \phn2.52$\pm$0.18 & \phn\phn7.1$\pm$1.1 & \null\hf $\times$ \hf  \pht    \hf \pht  \hf  \pht    \hf\null & 1,5,13 \\
N3877   & $-$20.60 & $-$21.73 & $-$22.29 & $-$23.76 & 76$\pm$1 &    335$\pm$\phn6 & F     \hf    L &    167$\pm$11    &    167$\pm$11    & Sc   & 17.12 & 0.52 & 3.2 &    \phn19.5$\pm$0.6 & \phn5.67$\pm$0.23 &    20.30$\pm$0.81 &    \phn35.6$\pm$2.4 & \null\hf $\times$ \hf $\times$ \hf   F   \hf $\times$ \hf\null &        \\
U6773   & $-$17.09 & $-$17.87 & $-$18.28 & $-$20.14 & 58$\pm$3 &    113$\pm$\phn5 & R     \hf    L & \phn45$\pm$\phn5 &    \nd           & Sm   & 19.48 & 0.28 & 3.3 & \phn\phn5.6$\pm$0.4 & \tlt0.20 \hf\null & \tlt0.62 \hf\null & \phd\tlt2.6         & \null\hf $\times$ \hf $\times$ \hf   R   \hf  \pht    \hf\null &        \\
N3893   & $-$20.55 & $-$21.45 & $-$21.86 & $-$23.56 & 49$\pm$2 &    382$\pm$12    & F/(D) \hf\null &    188$\pm$11    &    188$\pm$11    & Sc   & 17.15 & 0.45 & 4.7 &    \phn69.9$\pm$0.5 &    12.40$\pm$1.12 &    37.80$\pm$1.89 &       137.4$\pm$2.9 & \null\hf $\times$ \hf $\times$ \hf \pht  \hf  \pht    \hf\null & 8      \\
N3896   & $-$17.69 & $-$18.45 & $-$18.92 & $-$20.01 & 48$\pm$3 & \phn83$\pm$\phn4 & confused       &    \nd           &    \nd           & Sm   & 18.81 & 0.19 & 6.4 & \phn\phn2.5$\pm$0.1 & confused          & confused          & \phd\tlt3.1         & \null\hf $\times$ \hf  \pht    \hf \pht  \hf  \pht    \hf\null & 8      \\
N3917   & $-$19.65 & $-$20.63 & $-$21.05 & $-$22.40 & 79$\pm$2 &    275$\pm$\phn3 & F     \hf\null &    135$\pm$\phn3 &    135$\pm$\phn3 & Scd  & 18.66 & 0.57 & 3.0 &    \phn24.9$\pm$0.6 & \phn0.61$\pm$0.05 & \phn3.16$\pm$0.22 & \phd\tlt7.2         & \null\hf $\times$ \hf $\times$ \hf   F   \hf $\times$ \hf\null &        \\
U6818   & $-$17.40 & $-$18.10 & $-$18.46 & $-$19.71 & 75$\pm$3 &    151$\pm$\phn3 & R/(F) \hf    L & \phn73$\pm$\phn5 &\phn 73$\pm$\phn5 & Sd   & 20.08 & 0.33 & 3.4 &    \phn13.9$\pm$0.2 & \tlt0.15 \hf\null & \tlt0.43 \hf\null & \phn\phn2.4$\pm$1.0 & \null\hf $\times$ \hf $\times$ \hf \pht  \hf  \pht    \hf\null & 10     \\
N3949   & $-$20.22 & $-$20.96 & $-$21.31 & $-$22.98 & 55$\pm$2 &    320$\pm$\phn8 & F     \hf    L &    164$\pm$\phn7 &    164$\pm$\phn7 & Sbc  & 17.08 & 0.32 & 3.7 &    \phn44.8$\pm$0.4 &    10.42$\pm$0.63 &    24.94$\pm$1.25 &       134.1$\pm$3.6 & \null\hf $\times$ \hf $\times$ \hf   F   \hf $\times$ \hf\null &        \\
N3953   & $-$21.05 & $-$22.20 & $-$22.74 & $-$24.41 & 62$\pm$1 &    446$\pm$\phn5 & F     \hf\null &    223$\pm$\phn5 &    223$\pm$\phn5 & SBbc & 17.22 & 0.71 & 4.4 &    \phn39.3$\pm$0.8 & \phn3.68$\pm$0.18 &    28.49$\pm$7.69 &    \phn50.9$\pm$2.5 & \null\hf $\times$ \hf $\times$ \hf   F   \hf  \pht    \hf\null &        \\
U6894   & $-$16.51 & $-$17.39 & $-$17.59 & $-$19.01 & 83$\pm$3 &    124$\pm$\phn1 & R     \hf\null & \phn63$\pm$\phn5 &    \nd           & Scd  & 20.35 & 0.28 & 3.1 & \phn\phn5.8$\pm$0.2 & \tlt0.16 \hf\null & \tlt0.44 \hf\null & \phd\tlt2.7         & \null\hf $\times$ \hf $\times$ \hf   R   \hf $\times$ \hf\null &        \\
N3972   & $-$19.08 & $-$20.05 & $-$20.48 & $-$22.08 & 77$\pm$1 &    264$\pm$\phn2 & R     \hf    L &    134$\pm$\phn5 &    \nd           & Sbc  & 17.90 & 0.36 & 3.0 &    \phn16.6$\pm$0.4 & \phn1.01$\pm$0.06 & \phn3.66$\pm$0.22 & \phd\tlt5.8         & \null\hf $\times$ \hf $\times$ \hf   R   \hf $\times$ \hf\null &        \\
U6917   & $-$18.63 & $-$19.49 & $-$19.84 & $-$21.10 & 56$\pm$2 &    224$\pm$\phn7 & R/F   \hf\null &    104$\pm$\phn4 &    104$\pm$\phn4 & SBd  & 19.83 & 0.54 & 3.4 &    \phn26.2$\pm$0.3 & \phn0.26$\pm$0.04 & \phn1.02$\pm$0.18 & \phd\tlt4.4         & \null\hf $\times$ \hf $\times$ \hf   F   \hf $\times$ \hf\null &        \\
N3985   & $-$18.39 & $-$19.30 & $-$19.69 & $-$21.19 & 51$\pm$3 &    180$\pm$\phn9 & R     \hf\null & \phn93$\pm$\phn7 &    \nd           & Sm   & 17.56 & 0.21 & 2.9 &    \phn15.7$\pm$0.6 & \phn1.42$\pm$0.09 & \phn3.42$\pm$0.24 & \phn\phn9.7$\pm$1.4 & \null\hf $\times$ \hf $\times$ \hf   R   \hf  \pht    \hf\null & 14     \\
U6923   & $-$17.84 & $-$18.67 & $-$19.20 & $-$20.36 & 65$\pm$2 &    160$\pm$\phn4 & R     \hf    L & \phn81$\pm$\phn5 &    \nd           & Sdm  & 18.80 & 0.24 & 5.0 &    \phn10.7$\pm$0.6 & \phn0.37$\pm$0.04 & \phn0.90$\pm$0.19 & \phd\tlt2.6         & \null\hf $\times$ \hf $\times$ \hf   R   \hf  \pht    \hf\null &        \\
N3992   & $-$21.18 & $-$22.28 & $-$22.80 & $-$24.21 & 56$\pm$2 &    547$\pm$13    & F/D   \hf\null &    272$\pm$\phn6 &    242$\pm$\phn5 & SBbc & 17.45 & 0.77 & 4.0 &    \phn74.6$\pm$1.5 & \phn1.12$\pm$0.07 &    10.35$\pm$0.52 &    \phn30.2$\pm$7.6 & \null\hf $\times$ \hf $\times$ \hf   D   \hf  \pht    \hf\null &        \\
N3990   & $-$17.89 & $-$19.31 & $-$20.02 & $-$21.82 & 63$\pm$3 &    \nd           & \nd            &    \nd           &    \nd           & S0   & 16.79 & 0.15 & 8.5 &         \nd         & \tlt0.17 \hf\null & \tlt0.32 \hf\null & \phd\tlt1.4         & \null\hf  \pht    \hf  \pht    \hf \pht  \hf  \pht    \hf\null & 3,5    \\
N4013   & $-$20.08 & $-$21.40 & $-$22.07 & $-$23.83 & 90$\pm$1 &    377$\pm$\phn1 & F/D   \hf\null &    195$\pm$\phn3 &    177$\pm$\phn6 & Sb   & 16.44 & 0.38 & 3.7 &    \phn41.5$\pm$0.2 & \phn5.70$\pm$0.34 &    20.13$\pm$1.81 &    \phn36.3$\pm$0.8 & \null\hf $\times$ \hf $\times$ \hf   D   \hf $\times$ \hf\null & 6      \\
N4010   & $-$19.30 & $-$20.17 & $-$20.55 & $-$22.31 & 89$\pm$1 &    254$\pm$\phn1 & (R)/F \hf    L &    128$\pm$\phn9 &    128$\pm$\phn9 & SBd  & 19.41 & 0.64 & 2.9 &    \phn38.2$\pm$0.3 & \phn1.68$\pm$0.15 & \phn6.76$\pm$0.34 &    \phn16.9$\pm$1.6 & \null\hf $\times$ \hf $\times$ \hf   F   \hf $\times$ \hf\null & 6      \\
U6969   & $-$16.56 & $-$17.28 & $-$17.48 & $-$18.80 & 76$\pm$2 &    117$\pm$\phn7 & R     \hf\null & \phn79$\pm$\phn5 &    \nd           & Sm   & 20.34 & 0.25 & 3.0 & \phn\phn6.1$\pm$0.5 & \tlt0.10 \hf\null & \tlt0.48 \hf\null & \phd\tlt3.8         & \null\hf $\times$ \hf $\times$ \hf   R   \hf  \pht    \hf\null &        \\
U6973   & $-$19.21 & $-$20.67 & $-$21.28 & $-$23.23 & 71$\pm$3 &    363$\pm$\phn7 & F/D   \hf\null &    173$\pm$10    &    173$\pm$10    & Sab  & 15.72 & 0.18 & 5.3 &    \phn22.9$\pm$0.2 &      \nd          &      \nd          &       127.5$\pm$2.1 & \null\hf $\times$ \hf $\times$ \hf \pht  \hf  \pht    \hf\null & 9      \\
U6983   & $-$18.58 & $-$19.31 & $-$19.61 & $-$20.87 & 49$\pm$1 &    222$\pm$\phn4 & F     \hf\null &    107$\pm$\phn7 &    107$\pm$\phn7 & SBcd & 19.87 & 0.49 & 3.8 &    \phn38.5$\pm$0.6 & \phn0.29$\pm$0.06 & \phn1.34$\pm$0.20 & \phd\tlt5.4         & \null\hf $\times$ \hf $\times$ \hf   F   \hf $\times$ \hf\null &        \\
N4026   & $-$19.74 & $-$21.16 & $-$21.82 & $-$23.71 & 84$\pm$3 &    385$\pm$11    & \nd            &    \nd           &    \nd           & S0   & 17.13 & 0.43 & 7.3 &         \nd         & \tlt0.12 \hf\null & \tlt0.52 \hf\null & \phd\tlt1.4         & \null\hf $\times$ \hf  \pht    \hf \pht  \hf  \pht    \hf\null & 3,5    \\
N4051   & $-$20.71 & $-$21.72 & $-$22.18 & $-$23.54 & 49$\pm$3 &    309$\pm$14    & R/F   \hf    L &    159$\pm$13    &    159$\pm$13    & SBbc & 17.18 & 0.50 & 4.0 &    \phn35.6$\pm$0.8 & \phn7.14$\pm$0.93 &    23.92$\pm$1.20 &    \phn26.5$\pm$2.6 & \null\hf $\times$ \hf $\times$ \hf \pht  \hf  \pht    \hf\null & 15     \\
N4085   & $-$19.12 & $-$20.11 & $-$20.57 & $-$22.27 & 82$\pm$2 &    247$\pm$\phn7 & R/F   \hf    L &    134$\pm$\phn6 &    134$\pm$\phn6 & Sc   & 17.36 & 0.29 & 3.2 &    \phn14.6$\pm$0.9 & \phn5.49$\pm$0.27 &    14.61$\pm$0.73 &    \phn44.1$\pm$1.3 & \null\hf $\times$ \hf $\times$ \hf   F   \hf $\times$ \hf\null &        \\
N4088   & $-$20.95 & $-$21.94 & $-$22.45 & $-$24.00 & 69$\pm$2 &    362$\pm$\phn5 & F/(D) \hf    L &    173$\pm$14    &    173$\pm$14    & Sbc  & 17.25 & 0.62 & 3.0 &       102.9$\pm$1.1 &    19.88$\pm$0.99 &    54.47$\pm$2.72 &       222.3$\pm$1.9 & \null\hf $\times$ \hf $\times$ \hf   F   \hf  \pht    \hf\null &        \\
U7089   & $-$18.11 & $-$18.96 & $-$19.26 & $-$20.30 & 80$\pm$3 &    138$\pm$\phn2 & R     \hf    L & \phn79$\pm$\phn7 &    \nd           & Sdm  & 20.53 & 0.57 & 3.5 &    \phn17.0$\pm$0.6 &      \nd          &      \nd          & \phd\tlt3.4         & \null\hf $\times$ \hf $\times$ \hf   R   \hf  \pht    \hf\null &        \\
N4100   & $-$20.51 & $-$21.49 & $-$21.97 & $-$23.48 & 73$\pm$2 &    386$\pm$\phn5 & F/D   \hf\null &    195$\pm$\phn7 &    164$\pm$13    & Sbc  & 17.11 & 0.47 & 2.9 &    \phn41.6$\pm$0.7 & \phn8.10$\pm$0.49 &    21.72$\pm$0.87 &    \phn54.3$\pm$1.7 & \null\hf $\times$ \hf $\times$ \hf   D   \hf $\times$ \hf\null &        \\
U7094   & $-$16.67 & $-$17.69 & $-$18.16 & $-$19.78 & 70$\pm$3 & \phn76$\pm$\phn3 & R     \hf    L & \phn35$\pm$\phn6 &    \nd           & Sdm  & 19.68 & 0.27 & 4.5 & \phn\phn2.9$\pm$0.2 &      \nd          &      \nd          & \phd\tlt2.6         & \null\hf $\times$ \hf $\times$ \hf   R   \hf  \pht    \hf\null &        \\
N4102   & $-$19.86 & $-$21.20 & $-$21.73 & $-$23.57 & 56$\pm$2 &    393$\pm$10    & F     \hf\null &    178$\pm$11    &    178$\pm$11    & SBab & 16.78 & 0.33 & 7.6 & \phn\phn8.0$\pm$0.2 &    46.90$\pm$1.88 &    69.74$\pm$2.79 &       276.0$\pm$1.5 & \null\hf $\times$ \hf $\times$ \hf   F   \hf  \pht    \hf\null & 16     \\
N4111   & $-$20.01 & $-$21.44 & $-$22.13 & $-$23.76 & 90$\pm$3 &    \nd           & \nd            &    \nd           &    \nd           & S0   & 17.25 & 0.40 & 7.8 &         \nd         &      \nd          &      \nd          & \phn\phn9.7$\pm$0.6 & \null\hf  \pht    \hf  \pht    \hf \pht  \hf  \pht    \hf\null & 3,5    \\
N4117   & $-$17.36 & $-$18.92 & $-$19.57 & $-$21.38 & 68$\pm$3 &    251$\pm$10    & \nd            &    \nd           &    \nd           & S0   & 18.01 & 0.20 & 6.1 & \phn\phn6.9$\pm$1.1 &      \nd          &      \nd          & \phn\phn3.7$\pm$1.2 & \null\hf $\times$ \hf  \pht    \hf \pht  \hf  \pht    \hf\null & 11     \\
U7129   & $-$17.36 & $-$18.65 & $-$19.23 &   \nd    & 48$\pm$3 & \phn96$\pm$\phn8 & \nd            &    \nd           &    \nd           & Sa   &  \nd  &  \nd & 4.4 &         \nd         &      \nd          &      \nd          & \phd\tlt1.4         & \null\hf $\times$ \hf  \pht    \hf \pht  \hf  \pht    \hf\null & 2,5    \\
N4138   & $-$19.50 & $-$20.93 & $-$21.50 & $-$23.22 & 53$\pm$3 &    374$\pm$16    & F/D   \hf\null &    195$\pm$\phn7 &    147$\pm$12    & Sa   & 16.48 & 0.26 & 5.3 &    \phn19.2$\pm$0.7 &      \nd          &      \nd          &    \phn16.7$\pm$4.6 & \null\hf $\times$ \hf $\times$ \hf   D   \hf  \pht    \hf\null &        \\
N4143   & $-$19.35 & $-$20.83 & $-$21.54 & $-$23.50 & 60$\pm$3 &    \nd           & \nd            &    \nd           &    \nd           & S0   & 15.89 & 0.23 & 8.0 &         \nd         &      \nd          &      \nd          &    \phn10.8$\pm$1.0 & \null\hf  \pht    \hf  \pht    \hf \pht  \hf  \pht    \hf\null & 4,5    \\
N4157   & $-$20.72 & $-$21.83 & $-$22.33 & $-$24.04 & 82$\pm$3 &    398$\pm$\phn4 & F/D   \hf\null &    201$\pm$\phn7 &    185$\pm$10    & Sb   & 16.77 & 0.48 & 3.9 &       107.4$\pm$1.6 &    12.01$\pm$0.72 &    45.43$\pm$2.27 &       179.6$\pm$2.3 & \null\hf $\times$ \hf $\times$ \hf   D   \hf $\times$ \hf\null &        \\
N4183   & $-$19.46 & $-$20.16 & $-$20.46 & $-$21.74 & 82$\pm$2 &    228$\pm$\phn2 & F/D   \hf    L &    115$\pm$\phn6 &    109$\pm$\phn4 & Scd  & 19.47 & 0.59 & 3.2 &    \phn48.9$\pm$0.7 &      \nd          &      \nd          & \phd\tlt5.8         & \null\hf $\times$ \hf $\times$ \hf   D   \hf $\times$ \hf\null &        \\
N4218   & $-$17.88 & $-$18.68 & $-$19.06 & $-$20.55 & 53$\pm$3 &    150$\pm$\phn9 & R     \hf\null & \phn73$\pm$\phn7 &    \nd           & Sm   & 17.06 & 0.12 & 3.2 & \phn\phn7.8$\pm$0.2 & \phn1.09$\pm$0.08 & \phn2.29$\pm$0.23 & \phn\phn6.3$\pm$0.8 & \null\hf $\times$ \hf $\times$ \hf   R   \hf  \pht    \hf\null &        \\
N4217   & $-$20.33 & $-$21.54 & $-$22.16 & $-$23.90 & 86$\pm$2 &    381$\pm$\phn5 & F/D   \hf\null &    191$\pm$\phn6 &    178$\pm$\phn5 & Sb   & 17.17 & 0.54 & 3.3 &    \phn33.8$\pm$0.7 & \phn8.88$\pm$0.62 &    35.35$\pm$2.47 &       115.6$\pm$2.2 & \null\hf $\times$ \hf $\times$ \hf   D   \hf $\times$ \hf\null &        \\
N4220   & $-$20.03 & $-$21.29 & $-$21.91 & $-$23.13 & 78$\pm$3 &    399$\pm$\phn5 & \nd            &    \nd           &    \nd           & Sa   & 16.55 & 0.29 & 4.2 & \phn\phn4.4$\pm$0.3 & \phn1.57$\pm$0.13 & \phn5.53$\pm$0.50 & \phd\tlt4.9         & \null\hf $\times$ \hf  \pht    \hf \pht  \hf  \pht    \hf\null & 11     \\
N4346   & $-$19.27 & $-$20.70 & $-$21.42 & $-$23.15 & 77$\pm$3 &    \nd           & \nd            &    \nd           &    \nd           & S0   & 17.13 & 0.33 & 8.2 &         \nd         &      \nd          &      \nd          & \phd\tlt1.4         & \null\hf  \pht    \hf  \pht    \hf \pht  \hf  \pht    \hf\null & 3,5    \\
N4389   & $-$19.05 & $-$20.21 & $-$20.63 & $-$22.27 & 50$\pm$4 &    212$\pm$12    & R     \hf\null &    110$\pm$\phn8 &    \nd           & SBbc & 17.07 & 0.27 & 3.6 & \phn\phn7.6$\pm$0.2 &      \nd          &      \nd          &    \phn23.3$\pm$1.2 & \null\hf $\times$ \hf $\times$ \hf \pht  \hf  \pht    \hf\null & 17     \\
\enddata

\end{deluxetable}

%Table~3
\begin{deluxetable}{rl}
\tabletypesize{\small}
\tablewidth{88mm}
\setlength{\tabcolsep}{0.8mm}
\tablecaption{Comments on Table~2}
\tablehead{
\colhead{No.}      &
\colhead{\phantom{: } Comments \hfill\null} }

\startdata
  1 & : HI data from \citet{ft81} \\
  2 & : optical redshift from \citet{fkghpbmte99} \\
  3 & : optical redshift from RC3 \\
  4 & : optical redshift from \citet{dggp95} \\
  5 & : 1.4GHz flux or upper limit from the NVSS. \\
  6 & : strongly warped \\
  7 & : interacting pair N3769/1135+48 \\
  8 & : interacting pair N3893/N3896 \\
  9 & : interacting with U6962 \\
 10 & : interacting with faint dwarf? \\
 11 & : too little HI to retrieve \Vmax or \Vflat \\
 12 & : anomalous; extreme warp, conspicuous dust lane \\
 13 & : Markarian 186 \\
 14 & : anomalous; extreme warp, optically disturbed \\
 15 & : strongly lopsided, Seyfert \\
 16 & : HI in absorption against nucleus, LINER \\
 17 & : bar dominated \\
 18 & : 5-sigma upper limits to the IRAS flux given the rms noise in \\
    & \phantom{ :} the median coadded local IRAS scan. \\
\enddata
\end{deluxetable}

%Table~4
\begin{deluxetable}{ll}
\tabletypesize{\small}
\tablewidth{88mm}
\tablecaption{Galaxies excluded from the \HIsample\ based on
the selection criteria by Bernstein et al.(1994).  Several galaxies are
excluded for more than one reason.}

\tablehead{
\colhead{Criteria} & 
\colhead{Excludes} }

\startdata
non-interacting        & N3769, N3893, U6973, U6818                 \\
Sb-Sd morphology       & N3718, N3729, N3870, U6973, N4102          \\
                       & U7129, N4138, N4220 \hfill (earlier types) \\
                       & U6399, U6773, N3985, U6923, U6969          \\
                       & U7089, U7094, N4218 \hfill (later types)   \\
no prominent bar       & N3726, N3729, N3769, N3953, N3992          \\
                       & N4010, N4051, N4102, N4389                 \\
smooth outer isophotes & N3718, N3893, U6973, U6818, N3985          \\
                       & N4051, N4088                               \\
steep HI profile edges & N3769, N3985, N4218                        \\
\enddata
\end{deluxetable}

%Table~5
\begin{deluxetable}{ccrrrrrrccccccrrr}
\tabletypesize{\footnotesize}
\tablewidth{184mm}
\setlength{\tabcolsep}{1.25mm}
\renewcommand{\arraystretch}{0.85}

\tablecaption{Results from inverse least-squares linear fits to the
various samples, taking errors of 5\% in \WRIi, \Vmax\ and \Vflat\
into account as well as uncertainties of 0.05 mag in the opticals
passbands and 0.08 mag in the \Kp\ passband. It is assumed that
$\sigma^2_{\mbox{\scriptsize depth}}$+$\sigma^2_{\mbox{\scriptsize intr}}$=0.}
\hf = \hfill
\newcommand{\tlt}{{\tiny$<$}}
\newcommand{\pht}{\hspace{3mm}}

\tablehead{
\colhead{}                                &
\colhead{N}                               &
\multicolumn{3}{c}{Zero point}            &
\multicolumn{3}{c}{Slope}                 &
\multicolumn{6}{c}{Scatter}               &
\multicolumn{3}{c}{$\chi^2_{\mbox{red}}$} \\
\colhead{}                                &     
\colhead{}                                &     
\multicolumn{3}{c}{}                      &
\multicolumn{3}{c}{}                      &
\multicolumn{3}{c}{rms}      &
\multicolumn{3}{c}{bi-weight}             &
\multicolumn{3}{c}{}                      \\ 
\colhead{}                                &
\colhead{}                                &
\colhead{W$_{\mbox{\scriptsize R,I}}^{\mbox{\scriptsize i}}$} &
\colhead{2V$_{\mbox{\scriptsize max}}$}                       &
\colhead{2V$_{\mbox{\scriptsize flat}}$}                      &
\colhead{W$_{\mbox{\scriptsize R,I}}^{\mbox{\scriptsize i}}$} &
\colhead{2V$_{\mbox{\scriptsize max}}$}                       &
\colhead{2V$_{\mbox{\scriptsize flat}}$}                      &
\colhead{W$_{\mbox{\scriptsize R,I}}^{\mbox{\scriptsize i}}$} &
\colhead{2V$_{\mbox{\scriptsize m}}$}                       &
\colhead{2V$_{\mbox{\scriptsize f}}$}                      &
\colhead{W$_{\mbox{\scriptsize R,I}}^{\mbox{\scriptsize i}}$} &
\colhead{2V$_{\mbox{\scriptsize m}}$}                       &
\colhead{2V$_{\mbox{\scriptsize f}}$}                      &
\colhead{W$_{\mbox{\scriptsize R,I}}^{\mbox{\scriptsize i}}$} &
\colhead{2V$_{\mbox{\scriptsize m}}$}                       &
\colhead{2V$_{\mbox{\scriptsize f}}$}                      \\
\colhead{}                                &
\colhead{}                                &
\colhead{mag}                             &
\colhead{mag}                             &
\colhead{mag}                             &
\colhead{}                                &
\colhead{}                                &
\colhead{}                                &
\colhead{mag}                             &
\colhead{mag}                             &
\colhead{mag}                             & 
\colhead{mag}                             &
\colhead{mag}                             &
\colhead{mag}                             &
\colhead{}                                &
\colhead{}                                &
\colhead{}                                \\
\colhead{(1)}  &
\colhead{(2)}  &
\colhead{(3a)}  &
\colhead{(3b)}  &
\colhead{(3c)}  &
\colhead{(4a)}  &
\colhead{(4b)}  &
\colhead{(4c)} &
\colhead{(5a)} &
\colhead{(5b)} &
\colhead{(5c)} &
\colhead{(6a)} &
\colhead{(6b)} &
\colhead{(6c)} &
\colhead{(7a)} &
\colhead{(7b)} &
\colhead{(7c)} }

\startdata

\multicolumn{17}{l}{\underline{{\em HI} -- sample}} \\
 B          & 45 & $-$2.91$\pm$0.33 &                  &               &  $-$6.8$\pm$0.1 &                 &                 & 0.61 &      &      & 0.56 &      &      & 15.8 &      &      \\
 R          & 45 & $-$3.15$\pm$0.33 &                  &               &  $-$7.1$\pm$0.1 &                 &                 & 0.56 &      &      & 0.52 &      &      & 12.3 &      &      \\
 I          & 45 & $-$2.73$\pm$0.34 &                  &               &  $-$7.5$\pm$0.1 &                 &                 & 0.61 &      &      & 0.53 &      &      & 13.1 &      &      \\
 K$^\prime$ & 44 & $-$2.98$\pm$0.33 &                  &               &  $-$8.0$\pm$0.1 &                 &                 & 0.59 &      &      & 0.55 &      &      &  9.8 &      &      \\

\multicolumn{17}{l}{\underline{{\em DE} -- sample}} \\
 B          & 16 &    0.40$\pm$0.80 &                  &               &  $-$8.2$\pm$0.3 &                 &                 & 0.28 &      &      & 0.29 &      &      &  2.5 &      &      \\
 R          & 16 &    0.90$\pm$0.82 &                  &               &  $-$8.8$\pm$0.4 &                 &                 & 0.24 &      &      & 0.25 &      &      &  1.6 &      &      \\
 I          & 16 &    2.49$\pm$0.83 &                  &               &  $-$9.6$\pm$0.4 &                 &                 & 0.26 &      &      & 0.26 &      &      &  1.6 &      &      \\
 K$^\prime$ & 16 &    3.45$\pm$0.95 &                  &               & $-$10.6$\pm$0.4 &                 &                 & 0.32 &      &      & 0.34 &      &      &  1.9 &      &      \\

\multicolumn{17}{l}{\underline{{\em SI} -- sample}} \\
 B          & 38 & $-$2.09$\pm$0.37 &                  &               &  $-$7.1$\pm$0.1 &                 &                 & 0.48 &      &      & 0.48 &      &      &  8.9 &      &      \\
 R          & 38 & $-$1.96$\pm$0.38 &                  &               &  $-$7.6$\pm$0.1 &                 &                 & 0.45 &      &      & 0.43 &      &      &  7.0 &      &      \\
 I          & 38 & $-$1.33$\pm$0.39 &                  &               &  $-$8.0$\pm$0.2 &                 &                 & 0.49 &      &      & 0.47 &      &      &  7.4 &      &      \\
 K$^\prime$ & 38 & $-$1.77$\pm$0.45 &                  &               &  $-$8.5$\pm$0.2 &                 &                 & 0.55 &      &      & 0.49 &      &      &  7.7 &      &      \\

\multicolumn{17}{l}{\underline{{\em RC/FDR} -- sample}} \\
 B          & 31 & $-$2.37$\pm$0.41 & $-$2.14$\pm$0.41 &               &  $-$7.0$\pm$0.1 &  $-$7.1$\pm$0.1 &                 & 0.45 & 0.55 &      & 0.45 & 0.54 &      &  8.0 & 12.0 &      \\
 R          & 31 & $-$2.19$\pm$0.42 & $-$1.95$\pm$0.43 &               &  $-$7.5$\pm$0.2 &  $-$7.6$\pm$0.2 &                 & 0.43 & 0.55 &      & 0.40 & 0.46 &      &  6.5 & 10.5 &      \\
 I          & 31 & $-$1.49$\pm$0.43 & $-$1.19$\pm$0.44 &               &  $-$8.0$\pm$0.2 &  $-$8.1$\pm$0.2 &                 & 0.48 & 0.61 &      & 0.45 & 0.43 &      &  7.3 & 11.6 &      \\
 K$^\prime$ & 31 & $-$1.92$\pm$0.48 & $-$1.62$\pm$0.49 &               &  $-$8.4$\pm$0.2 &  $-$8.5$\pm$0.2 &                 & 0.56 & 0.71 &      & 0.48 & 0.44 &      &  8.1 & 12.8 &      \\

\multicolumn{17}{l}{\underline{{\em RC/FD} -- sample}} \\
 B          & 22 &    0.37$\pm$0.74 &    0.36$\pm$0.74 & \phantom{$-$}1.66$\pm$0.79 &  $-$8.1$\pm$0.3 &  $-$8.1$\pm$0.3 &  $-$8.7$\pm$0.3 & 0.44 & 0.47 & 0.39 & 0.46 & 0.48 & 0.41 &  6.1 &  6.9 &  4.2 \\
 R          & 22 &    0.50$\pm$0.75 &    0.41$\pm$0.75 & 2.01$\pm$0.81 &  $-$8.6$\pm$0.3 &  $-$8.5$\pm$0.3 &  $-$9.2$\pm$0.4 & 0.35 & 0.37 & 0.31 & 0.37 & 0.38 & 0.33 &  3.4 &  3.9 &  2.3 \\
 I          & 22 &    1.74$\pm$0.76 &    1.59$\pm$0.76 & 3.37$\pm$0.83 &  $-$9.2$\pm$0.3 &  $-$9.2$\pm$0.3 & $-$10.0$\pm$0.4 & 0.36 & 0.37 & 0.31 & 0.37 & 0.37 & 0.33 &  3.1 &  3.4 &  2.1 \\
 K$^\prime$ & 22 &    2.25$\pm$0.84 &    1.94$\pm$0.83 & 3.96$\pm$0.93 & $-$10.1$\pm$0.4 & $-$10.0$\pm$0.4 & $-$10.8$\pm$0.4 & 0.40 & 0.38 & 0.32 & 0.40 & 0.33 & 0.32 &  3.1 &  2.8 &  1.8 \\

\multicolumn{17}{l}{\underline{{\em RC/F} -- sample}} \\
 B          & 15 &    2.55$\pm$0.94 &                  & 2.85$\pm$0.96 &  $-$9.0$\pm$0.4 &                 &  $-$9.2$\pm$0.4 & 0.38 &      & 0.39 & 0.33 &      & 0.42 &  3.8 &      &  3.8 \\
 R          & 15 &    2.81$\pm$0.96 &                  & 3.09$\pm$0.98 &  $-$9.6$\pm$0.4 &                 &  $-$9.7$\pm$0.5 & 0.32 &      & 0.31 & 0.32 &      & 0.35 &  2.4 &      &  2.3 \\
 I          & 15 &    4.09$\pm$0.99 &                  & 4.33$\pm$1.00 & $-$10.2$\pm$0.5 &                 & $-$10.4$\pm$0.5 & 0.33 &      & 0.31 & 0.33 &      & 0.35 &  2.3 &      &  2.0 \\
 K$^\prime$ & 15 &    4.68$\pm$1.09 &                  & 4.74$\pm$1.09 & $-$11.1$\pm$0.5 &                 & $-$11.2$\pm$0.5 & 0.35 &      & 0.26 & 0.40 &      & 0.28 &  2.0 &      &  1.1 \\

\multicolumn{17}{l}{\underline{{\em RC/FD} -- sample, excl. N3992}} \\
 B          & 21 &    1.35$\pm$0.82 &    1.34$\pm$0.82 & 2.46$\pm$0.87 &  $-$8.5$\pm$0.4 &  $-$8.5$\pm$0.4 &  $-$9.0$\pm$0.4 & 0.45 & 0.48 & 0.40 & 0.45 & 0.49 & 0.41 &  5.7 &  6.5 &  4.0 \\
 R          & 21 &    1.45$\pm$0.82 &    1.34$\pm$0.82 & 2.81$\pm$0.88 &  $-$9.0$\pm$0.4 &  $-$8.9$\pm$0.4 &  $-$9.6$\pm$0.4 & 0.33 & 0.36 & 0.30 & 0.35 & 0.38 & 0.32 &  2.9 &  3.4 &  2.1 \\
 I          & 21 &    2.79$\pm$0.83 &    2.60$\pm$0.83 & 4.27$\pm$0.89 &  $-$9.7$\pm$0.4 &  $-$9.6$\pm$0.4 & $-$10.4$\pm$0.4 & 0.33 & 0.35 & 0.30 & 0.35 & 0.35 & 0.32 &  2.5 &  2.7 &  1.7 \\
 K$^\prime$ & 21 &    3.58$\pm$0.93 &    3.19$\pm$0.92 & 5.12$\pm$1.00 & $-$10.6$\pm$0.4 & $-$10.5$\pm$0.4 & $-$11.3$\pm$0.5 & 0.34 & 0.31 & 0.26 & 0.35 & 0.31 & 0.27 &  2.0 &  1.7 &  1.1 \\

%\multicolumn{17}{l}{\underline{{\em RC/F} -- sample}} \\
% B          & 15 &    1.70$\pm$0.36 &    2.84$\pm$1.09 & 2.84$\pm$1.09 &  $-$8.7$\pm$0.1 &  $-$9.2$\pm$0.4 &  $-$9.2$\pm$0.4 & 0.37 & 0.39 & 0.39 & 0.35 & 0.42 & 0.42 & 16.5 &  3.8 &  3.8 \\
% R          & 15 &    1.68$\pm$0.37 &    3.10$\pm$1.09 & 3.10$\pm$1.09 &  $-$9.1$\pm$0.1 &  $-$9.7$\pm$0.5 &  $-$9.7$\pm$0.5 & 0.31 & 0.31 & 0.31 & 0.32 & 0.35 & 0.35 & 14.0 &  2.3 &  2.3 \\
% I          & 15 &    2.73$\pm$0.38 &    4.34$\pm$1.17 & 4.34$\pm$1.17 &  $-$9.7$\pm$0.2 & $-$10.4$\pm$0.5 & $-$10.4$\pm$0.5 & 0.33 & 0.31 & 0.31 & 0.33 & 0.35 & 0.35 & 14.6 &  2.0 &  2.0 \\
% K$^\prime$ & 15 &    2.56$\pm$0.50 &    4.72$\pm$1.31 & 4.72$\pm$1.31 & $-$10.3$\pm$0.2 & $-$11.1$\pm$0.5 & $-$11.1$\pm$0.5 & 0.35 & 0.26 & 0.26 & 0.36 & 0.28 & 0.28 &  9.2 &  1.1 &  1.1 \\

\enddata
\end{deluxetable}

%Table~6
\begin{deluxetable}{lcllcccllc}
\tabletypesize{\scriptsize}
\tablewidth{88mm}
\setlength{\tabcolsep}{1.0mm}
\renewcommand{\arraystretch}{1.}
\newcommand{\phx}{\phantom{$-$}}

\tablecaption{Values of \Xred from least-squares fits to the residuals,
excluding N3992 and assuming \sigdepth=0.}

\tablehead{
\colhead{Param.}             &
\multicolumn{4}{c}{B-band}   &
\colhead{}                   &
\multicolumn{4}{c}{\Kp-band} \\
\colhead{}                   &
\colhead{N}                  &
\colhead{slope}              &
\colhead{}                   &
\colhead{\Xred}              & 
\colhead{}                   &
\colhead{N}                  & 
\colhead{slope}              &
\colhead{}                   &
\colhead{\Xred}              }

\startdata
\Vsys                                                     & 21 &  $-$0.00063 & $\pm$ 0.00037 & 3.9 & & 21 &  $-$0.00039 & $\pm$ 0.00047 & 1.0 \\
SGL                                                       & 21 &\phx 0.014   & $\pm$ 0.013   & 4.0 & & 21 &\phx 0.038   & $\pm$ 0.017   & 0.8 \\
Type                                                      & 20 &\phx 0.15    & $\pm$ 0.03    & 2.4 & & 21 &\phx 0.003   & $\pm$ 0.029   & 1.1 \\
cos(i)                                                    & 21 &  $-$0.08    & $\pm$ 0.19    & 4.0 & & 21 &  $-$0.55    & $\pm$ 0.24    & 0.8 \\
M$_{\mbox{\scriptsize HI}}$/L$_{\mbox{\Kp}}$              & 21 &\phx 0.47    & $\pm$ 0.08    & 2.4 & & 21 &\phx 0.03    & $\pm$ 0.10    & 1.1 \\
L$_{\mbox{\scriptsize FIR}}$/L$_{\mbox{\Kp}}$             & 16 &  $-$0.32    & $\pm$ 0.17    & 3.2 & & 16 &\phx 0.00    & $\pm$ 0.22    & 0.9 \\
B-I                                                       & 21 &  $-$0.80    & $\pm$ 0.13    & 2.2 & & 21 &  $-$0.09    & $\pm$ 0.17    & 1.0 \\
$\mu_{\mbox{\scriptsize 0,\Kp}}^{\mbox{\scriptsize b,i}}$ & 21 &\phx 0.13    & $\pm$ 0.03    & 3.3 & & 21 &\phx 0.016   & $\pm$ 0.041   & 1.1 \\
\enddata
\end{deluxetable}

%Table~7
\begin{deluxetable}{lccccccccc}
\tabletypesize{\footnotesize}
\tablewidth{184mm}
\setlength{\tabcolsep}{2.6mm}
\renewcommand{\arraystretch}{0.70}

\tablecaption{95\% confidence intervals for
$\sqrt{\sigma^2_{\mbox{\scriptsize
depth}}+\sigma^2_{\mbox{\scriptsize intr}}}$ and for \sigintr\ in
case \sigdepth=0.17. The numbers for each entry refer to the
minimum, the most likely and the maximum intrinsic scatter.}

\hf = \hfill
% \nd = \nodata
\newcommand{\tlt}{{\tiny$<$}}
\newcommand{\pht}{\hspace{3mm}}

\tablehead{
\colhead{}                                &
\colhead{}                                &
\colhead{}                                &
\multicolumn{3}{c}{$\sqrt{\sigma^2_{\mbox{\scriptsize depth}}+\sigma^2_{\mbox{\scriptsize intr}}}$} &
\colhead{}                                &
\multicolumn{3}{c}{\sigintr (\sigdepth=0.17)} \\
\colhead{}                                &
\colhead{}                                &
\colhead{}                                &
\colhead{W$_{\mbox{\scriptsize R,I}}^{\mbox{\scriptsize i}}$} &
\colhead{2V$_{\mbox{\scriptsize max}}$}                       &
\colhead{2V$_{\mbox{\scriptsize flat}}$}                      &
\colhead{}                                &
\colhead{W$_{\mbox{\scriptsize R,I}}^{\mbox{\scriptsize i}}$} &
\colhead{2V$_{\mbox{\scriptsize max}}$}                       &
\colhead{2V$_{\mbox{\scriptsize flat}}$}                      \\
\colhead{}                                &
\colhead{}                                &
\colhead{}                                &
\colhead{mag}                             &
\colhead{mag}                             &
\colhead{mag}                             &
\colhead{}                                &
\colhead{mag}                             &
\colhead{mag}                             &
\colhead{mag}                             }

\startdata

\noalign{\vspace{0.8mm}}
\multicolumn{10}{l}{\underline{{\em HI} -- sample}} \\
\noalign{\vspace{0.8mm}}
N=45 & B         && 0.48 -- 0.57 -- 0.70 &                      &                      && 0.45 -- 0.54 -- 0.68 &                      &                      \\
     & R         && 0.44 -- 0.53 -- 0.65 &                      &                      && 0.41 -- 0.50 -- 0.63 &                      &                      \\
     & I         && 0.47 -- 0.57 -- 0.70 &                      &                      && 0.44 -- 0.54 -- 0.68 &                      &                      \\
     & K$^\prime$&& 0.46 -- 0.56 -- 0.69 &                      &                      && 0.43 -- 0.53 -- 0.67 &                      &                      \\
\noalign{\vspace{0.8mm}}
\multicolumn{10}{l}{\underline{{\em DE} -- sample}} \\
\noalign{\vspace{0.8mm}}
N=16 & B         && 0.14 -- 0.24 -- 0.39 &                      &                      && 0.00 -- 0.17 -- 0.35 &                      &                      \\
     & R         && 0.04 -- 0.18 -- 0.32 &                      &                      && 0.00 -- 0.06 -- 0.27 &                      &                      \\
     & I         && 0.00 -- 0.18 -- 0.34 &                      &                      && 0.00 -- 0.06 -- 0.29 &                      &                      \\
     & K$^\prime$&& 0.11 -- 0.26 -- 0.44 &                      &                      && 0.00 -- 0.20 -- 0.41 &                      &                      \\
\noalign{\vspace{0.8mm}}
\multicolumn{10}{l}{\underline{{\em SI} -- sample}} \\
\noalign{\vspace{0.8mm}}
N=38 & B         && 0.36 -- 0.45 -- 0.56 &                      &                      && 0.32 -- 0.42 -- 0.53 &                      &                      \\
     & R         && 0.34 -- 0.42 -- 0.53 &                      &                      && 0.29 -- 0.38 -- 0.50 &                      &                      \\
     & I         && 0.37 -- 0.46 -- 0.57 &                      &                      && 0.33 -- 0.43 -- 0.54 &                      &                      \\
     & K$^\prime$&& 0.42 -- 0.52 -- 0.65 &                      &                      && 0.38 -- 0.49 -- 0.63 &                      &                      \\
\noalign{\vspace{0.8mm}}
\multicolumn{10}{l}{\underline{{\em RC/FDR} -- sample}} \\
\noalign{\vspace{0.8mm}}
N=31 & B         && 0.33 -- 0.42 -- 0.55 & 0.42 -- 0.52 -- 0.67 &                      && 0.28 -- 0.38 -- 0.52 & 0.38 -- 0.49 -- 0.65 &                      \\
     & R         && 0.32 -- 0.40 -- 0.53 & 0.42 -- 0.52 -- 0.67 &                      && 0.27 -- 0.36 -- 0.50 & 0.35 -- 0.49 -- 0.65 &                      \\
     & I         && 0.35 -- 0.45 -- 0.59 & 0.46 -- 0.55 -- 0.75 &                      && 0.32 -- 0.42 -- 0.56 & 0.43 -- 0.55 -- 0.73 &                      \\
     & K$^\prime$&& 0.42 -- 0.53 -- 0.69 & 0.54 -- 0.68 -- 0.87 &                      && 0.38 -- 0.50 -- 0.67 & 0.51 -- 0.66 -- 0.85 &                      \\
\noalign{\vspace{0.8mm}}
\multicolumn{10}{l}{\underline{{\em RC/FD} -- sample}} \\
\noalign{\vspace{0.8mm}}
N=22 & B         && 0.30 -- 0.41 -- 0.56 & 0.33 -- 0.43 -- 0.60 & 0.25 -- 0.35 -- 0.49 && 0.25 -- 0.37 -- 0.53 & 0.28 -- 0.39 -- 0.58 & 0.18 -- 0.31 -- 0.46 \\
     & R         && 0.21 -- 0.31 -- 0.44 & 0.24 -- 0.33 -- 0.47 & 0.15 -- 0.25 -- 0.38 && 0.12 -- 0.26 -- 0.41 & 0.14 -- 0.28 -- 0.44 & 0.00 -- 0.18 -- 0.34 \\
     & I         && 0.21 -- 0.31 -- 0.45 & 0.22 -- 0.33 -- 0.47 & 0.14 -- 0.25 -- 0.38 && 0.12 -- 0.26 -- 0.42 & 0.17 -- 0.28 -- 0.44 & 0.00 -- 0.18 -- 0.34 \\
     & K$^\prime$&& 0.24 -- 0.35 -- 0.51 & 0.21 -- 0.32 -- 0.47 & 0.11 -- 0.24 -- 0.38 && 0.17 -- 0.31 -- 0.48 & 0.12 -- 0.27 -- 0.44 & 0.00 -- 0.17 -- 0.34 \\
\noalign{\vspace{0.8mm}}
\multicolumn{10}{l}{\underline{{\em RC/F} -- sample}} \\
\noalign{\vspace{0.8mm}}
N=15 & B         && 0.23 -- 0.36 -- 0.55 &                      & 0.23 -- 0.36 -- 0.56 && 0.15 -- 0.32 -- 0.52 &                      & 0.15 -- 0.32 -- 0.53 \\
     & R         && 0.15 -- 0.27 -- 0.45 &                      & 0.14 -- 0.27 -- 0.44 && 0.00 -- 0.21 -- 0.42 &                      & 0.00 -- 0.21 -- 0.41 \\
     & I         && 0.15 -- 0.28 -- 0.47 &                      & 0.11 -- 0.25 -- 0.43 && 0.00 -- 0.22 -- 0.44 &                      & 0.00 -- 0.18 -- 0.39 \\
     & K$^\prime$&& 0.13 -- 0.28 -- 0.49 &                      & 0.00 -- 0.14 -- 0.33 && 0.00 -- 0.22 -- 0.46 &                      & 0.00 -- 0.00 -- 0.28 \\
\noalign{\vspace{0.8mm}}
\multicolumn{10}{l}{\underline{{\em RC/FD} -- sample, excl. N3992}} \\
\noalign{\vspace{0.8mm}}
N=21 & B         && 0.30 -- 0.41 -- 0.57 & 0.33 -- 0.44 -- 0.61 & 0.25 -- 0.35 -- 0.50 && 0.25 -- 0.37 -- 0.54 & 0.28 -- 0.41 -- 0.59 & 0.18 -- 0.31 -- 0.47 \\
     & R         && 0.19 -- 0.29 -- 0.42 & 0.22 -- 0.32 -- 0.46 & 0.13 -- 0.23 -- 0.37 && 0.08 -- 0.23 -- 0.38 & 0.14 -- 0.27 -- 0.43 & 0.00 -- 0.15 -- 0.33 \\
     & I         && 0.17 -- 0.27 -- 0.41 & 0.19 -- 0.29 -- 0.44 & 0.08 -- 0.21 -- 0.35 && 0.00 -- 0.21 -- 0.37 & 0.08 -- 0.23 -- 0.41 & 0.00 -- 0.12 -- 0.31 \\
     & K$^\prime$&& 0.14 -- 0.26 -- 0.41 & 0.09 -- 0.22 -- 0.37 & 0.00 -- 0.10 -- 0.27 && 0.00 -- 0.20 -- 0.37 & 0.00 -- 0.14 -- 0.33 & 0.00 -- 0.00 -- 0.21 \\

\enddata
\end{deluxetable}

%Table~8
\begin{deluxetable}{lcccccccccccccccc}
\tabletypesize{\footnotesize}
\tablewidth{184mm}
\setlength{\tabcolsep}{1.45mm}
\renewcommand{\arraystretch}{0.95}

\tablecaption{Various properties of the HSB/LSB pair N3917 and N4085.}
\newcommand{\tlt}{{\tiny$<$}}
\newcommand{\pht}{\hspace{3mm}}
\newcommand{\php}{\phantom{$-$}}

\tablehead{
\colhead{Name}                                &
\colhead{Type}                                &
\colhead{$\mu^i_0${\scriptsize (K$^\prime$)}} &
\colhead{V$_{\mbox{\scriptsize flat}}$}       & 
\colhead{M$^{\mbox{\scriptsize b,i}}_{\mbox{\scriptsize B}}$}          &
\colhead{M$^{\mbox{\scriptsize b,i}}_{\mbox{\scriptsize K$^\prime$}}$} &
\colhead{$\Delta$M$^{\mbox{\scriptsize b,i}}_{\mbox{\scriptsize B}}$}          &
\colhead{$\Delta$M$^{\mbox{\scriptsize b,i}}_{\mbox{\scriptsize K$^\prime$}}$} &
\colhead{A$^i_{\mbox{\scriptsize B}}$}                                 &
\colhead{A$^i_{\mbox{\scriptsize \Kp}}$}                               &
\colhead{M$^{\mbox{\scriptsize b}}_{\mbox{\scriptsize B}}$}            &
\colhead{M$^{\mbox{\scriptsize b}}_{\mbox{\scriptsize \Kp}}$}          &
\colhead{F$_{\mbox{\scriptsize  60$\mu$m}}$}        &
\colhead{F$_{\mbox{\scriptsize 100$\mu$m}}$}        &
\colhead{L$_{\mbox{\tiny FIR}}$}        &
\colhead{$\frac{L_{\mbox{\tiny FIR}}} {L^b_{\mbox{\tiny B}}}$}    &
\colhead{$\frac{L_{\mbox{\tiny FIR}}} {L^b_{\mbox{\tiny \Kp}}}$}  \\
\colhead{}                                    &
\colhead{}                                    &
\colhead{mag/"}                               &
\colhead{km/s}                                &
\colhead{mag}                                 &
\colhead{mag}                                 &
\colhead{mag}                                 &
\colhead{mag}                                 &
\colhead{mag}                                 &
\colhead{mag}                                 &
\colhead{mag}                                 &
\colhead{mag}                                 &
\colhead{Jy}                                  &
\colhead{Jy}                                  &
\colhead{10$^9$L$_\odot$}                     &
\colhead{}                                    &
\colhead{}                                    \\
\colhead{(1)}  &
\colhead{(2)}  &
\colhead{(3)}  &
\colhead{(4)}  &
\colhead{(5)}  &
\colhead{(6)}  &
\colhead{(7)}  &
\colhead{(8)}  &
\colhead{(9)}  &
\colhead{(10)} &
\colhead{(11)} &
\colhead{(12)} &
\colhead{(13)} &
\colhead{(14)} &
\colhead{(15)} &
\colhead{(16)} &
\colhead{(17)} }

\startdata

N3917 & Scd & 18.66 & 135$\pm$3 & $-$19.65 & $-$22.40 & \php0.23 & \php0.05 & 0.87 & 0.12 & $-$18.78 & $-$22.28 & 0.61$\pm$0.05 & \phn3.16$\pm$0.22 & 0.64 & 0.13 & 0.04 \\
N4085 & Sc  & 17.36 & 134$\pm$6 & $-$19.12 & $-$22.27 &  $-$0.27 &  $-$0.05 & 0.78 & 0.11 & $-$18.34 & $-$22.16 & 5.49$\pm$0.27 &    14.61$\pm$0.73 & 3.91 & 1.18 & 0.25 \\

\enddata

\end{deluxetable}

%Table~9
\begin{deluxetable}{cccccc}
\tabletypesize{\footnotesize}
\tablewidth{88mm}
\setlength{\tabcolsep}{3.0mm}
\renewcommand{\arraystretch}{0.95}

\tablecaption{Statistical properties of the baryonic TF-relation for
various values of the mass-to-light ratio of the gas. It is assumed that 
$\sigma^2_{\mbox{\scriptsize depth}}$+$\sigma^2_{\mbox{\scriptsize intr}}$=0.}
\newcommand{\tlt}{{\tiny$<$}}
\newcommand{\php}{\phantom{$-$}}

\tablehead{
\colhead{({\em M}$_{\mbox{\scriptsize gas}}$/L$_{\mbox{\scriptsize \Kp}}$)} &
\colhead{zero point} &
\colhead{slope}      &
\colhead{rms}        & 
\colhead{\Xred}      &
\colhead{Q}          \\
\colhead{}     &
\colhead{mag}  &
\colhead{}     &
\colhead{mag}  &
\colhead{}     &
\colhead{}     \\
\colhead{(1)}  &
\colhead{(2)}  &
\colhead{(3)}  &
\colhead{(4)}  &
\colhead{(5)}  &
\colhead{(6)}  }

\startdata

 0.1 &  $-$5.63 & \phn$-$7.4 & 0.53 & 9.2 & 0.00 \\
 0.2 &  $-$4.38 & \phn$-$7.8 & 0.41 & 5.1 & 0.00 \\
 0.4 &  $-$2.36 & \phn$-$8.5 & 0.33 & 2.9 & 0.00 \\
 0.8 &  $-$0.17 & \phn$-$9.3 & 0.28 & 1.7 & 0.02 \\
 1.6 & \php1.66 &    $-$10.0 & 0.25 & 1.2 & 0.17 \\
 3.2 & \php3.07 &    $-$10.5 & 0.24 & 1.1 & 0.33 \\
 6.4 & \php3.93 &    $-$10.9 & 0.25 & 1.0 & 0.38 \\
12.8 & \php4.47 &    $-$11.1 & 0.25 & 1.0 & 0.37 \\

\enddata

\end{deluxetable}

\clearpage

\begin{figure}
\epsscale{0.48}
\plotone{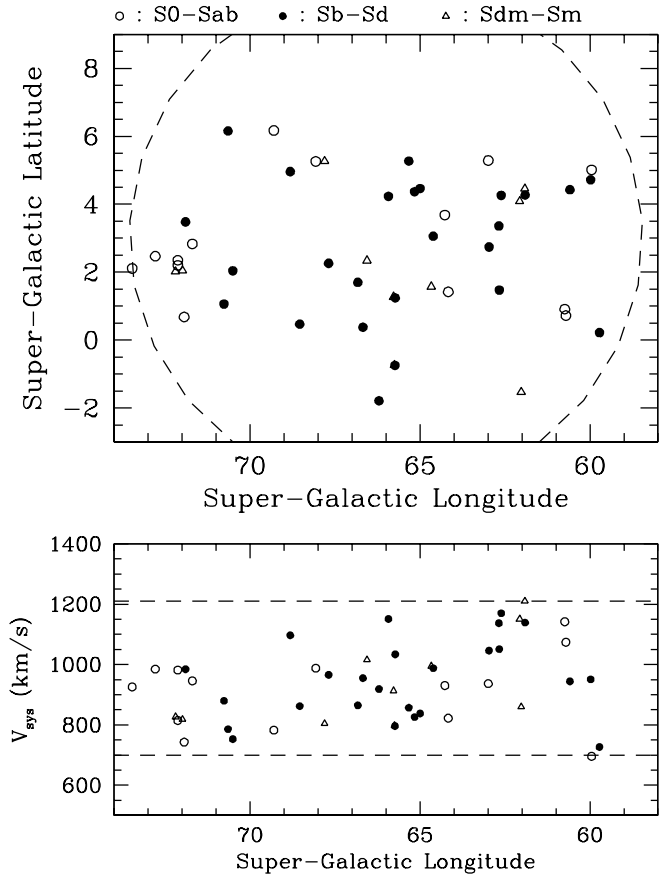}
\caption{Distribution on the sky and in redshift of all 49 galaxies in
the cluster brighter than M$_{\mbox{\scriptsize B}}=-16.8$ and more
inclined than 45$^{\circ}$ (the {\em complete} sample). The dashed lines
indicate the boundaries of the spatial and velocity window in which
the galaxies were selected.}
\end{figure}

\begin{figure*}
\epsscale{1.00}
\plotone{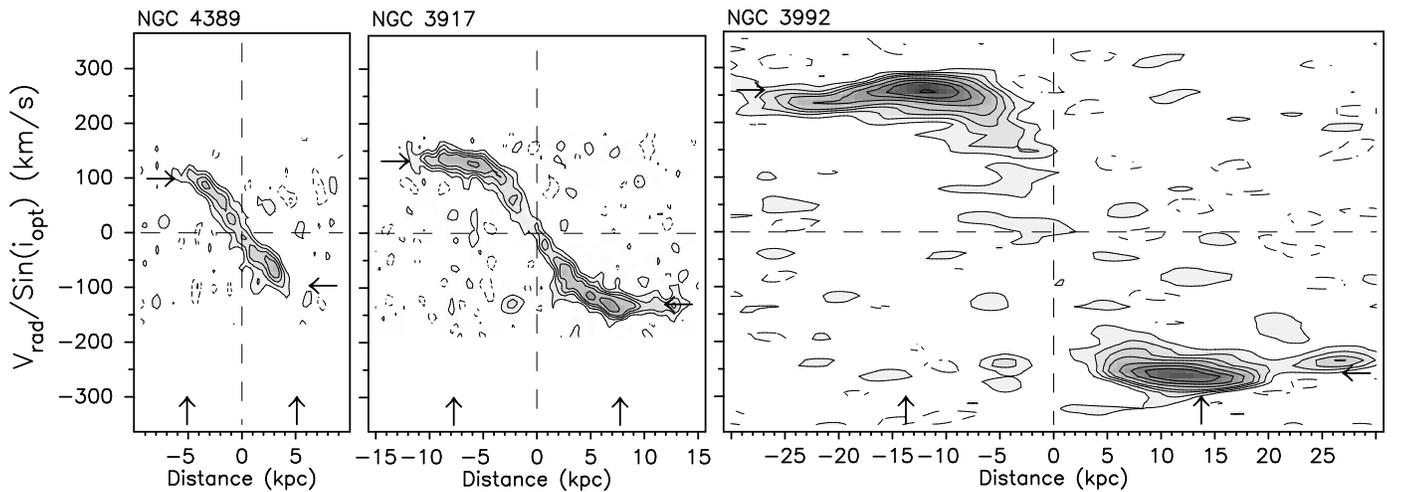}
\caption{Proto-type examples of the three catagories of rotation
curves.  {\bf Left}: Galaxy with a rotation curve that rises
continuously until the last measured point. The measured maximum
rotational velocity \Vmax\ is set by the extent of the HI disk
(R-curve). {\bf Middle}: The `classical' rotation curve; a gentle rise
in the central regions with a smooth transition into the extended flat
part (F-curve). {\bf Right}: A rotation curve that reaches a maximum
in the optical regions after which it declines somewhat to an extended
flat part in the outer disk. In this case, the maximum rotation
velocity exceeds the amplitude of the flat part (D-curve). The
vertical arrows indicate $\pm$R$_{25}$ and the horizontal arrows
indicate the rotational velocities as inferred from the global
profiles.}
\end{figure*}

\begin{figure*}
\epsscale{1.00}
\plotone{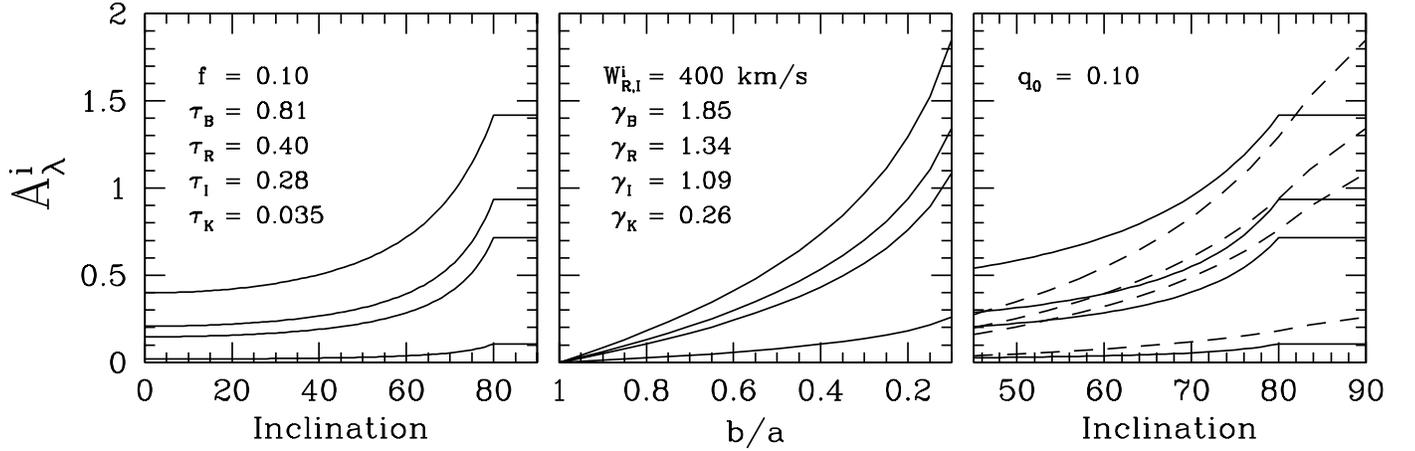}
\caption{Comparison of the two most commonly used correction methods
for internal extinction. Left panel: the TFq model. Middle panel: the
$\gamma$Log($a$/$b$) scheme. Right panel: the two models
overlayed. See Section~6 for details.}
\end{figure*}

\begin{figure*}
\epsscale{1.00}
\plotone{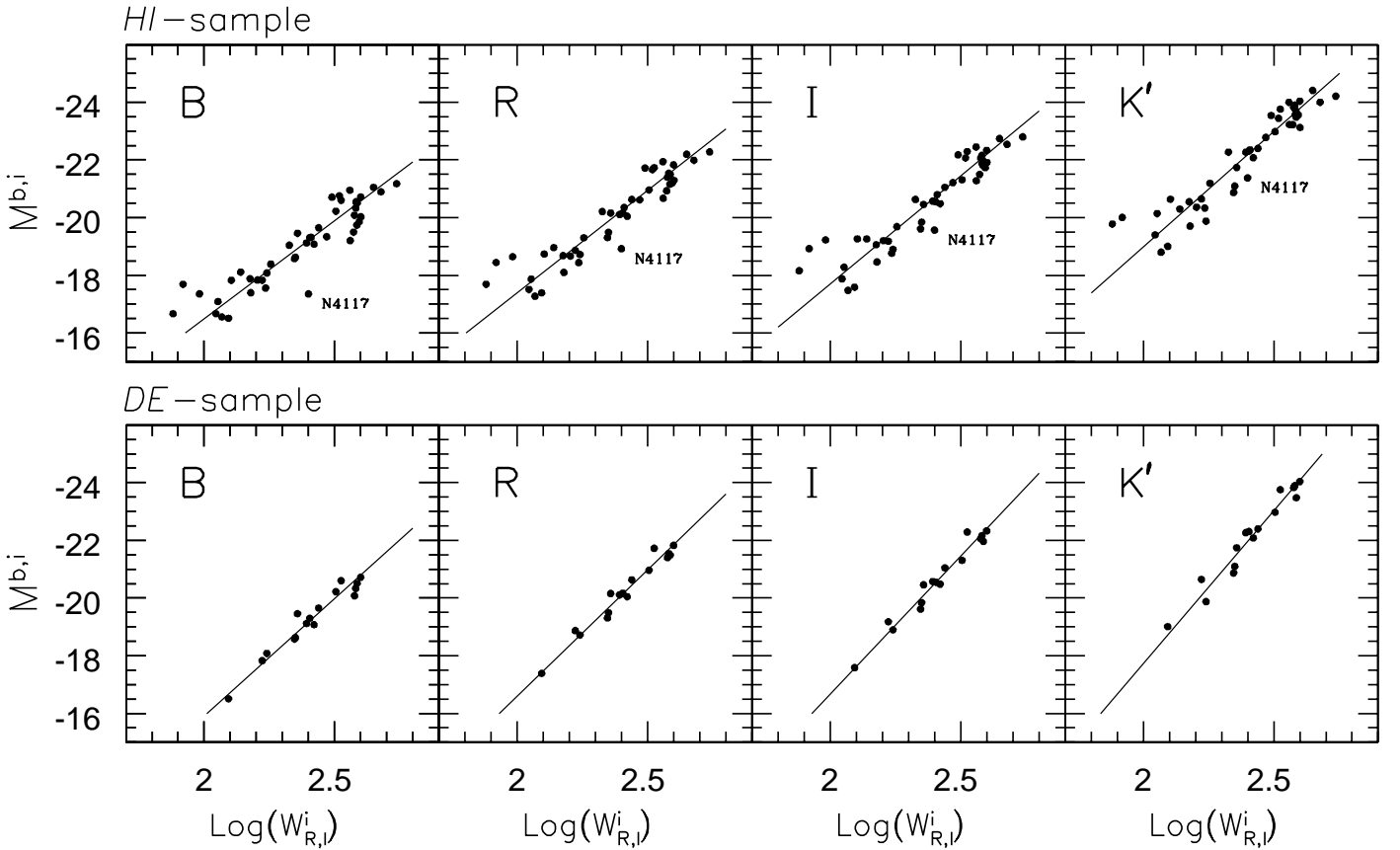}
\caption{TF-relations using the corrected widths \WRIi\ of the global
HI profiles.  {\bf Upper panels}: TF-relations for all 45 galaxies in
the \HIsample.  {\bf Lower panels}: TF-relations for the 16 galaxies
in the strictly selected \DEsample.  Solid lines show the results of
the inverse least-squares fits to these particular subsamples. Note
that the \DEsample\ displays steeper and tighter correlations with
less scatter, illustrating the merit of imposing selection criteria
when using the TF-relation as a distance estimator. }
\end{figure*}

\begin{figure*}
\epsscale{1.00}
\plotone{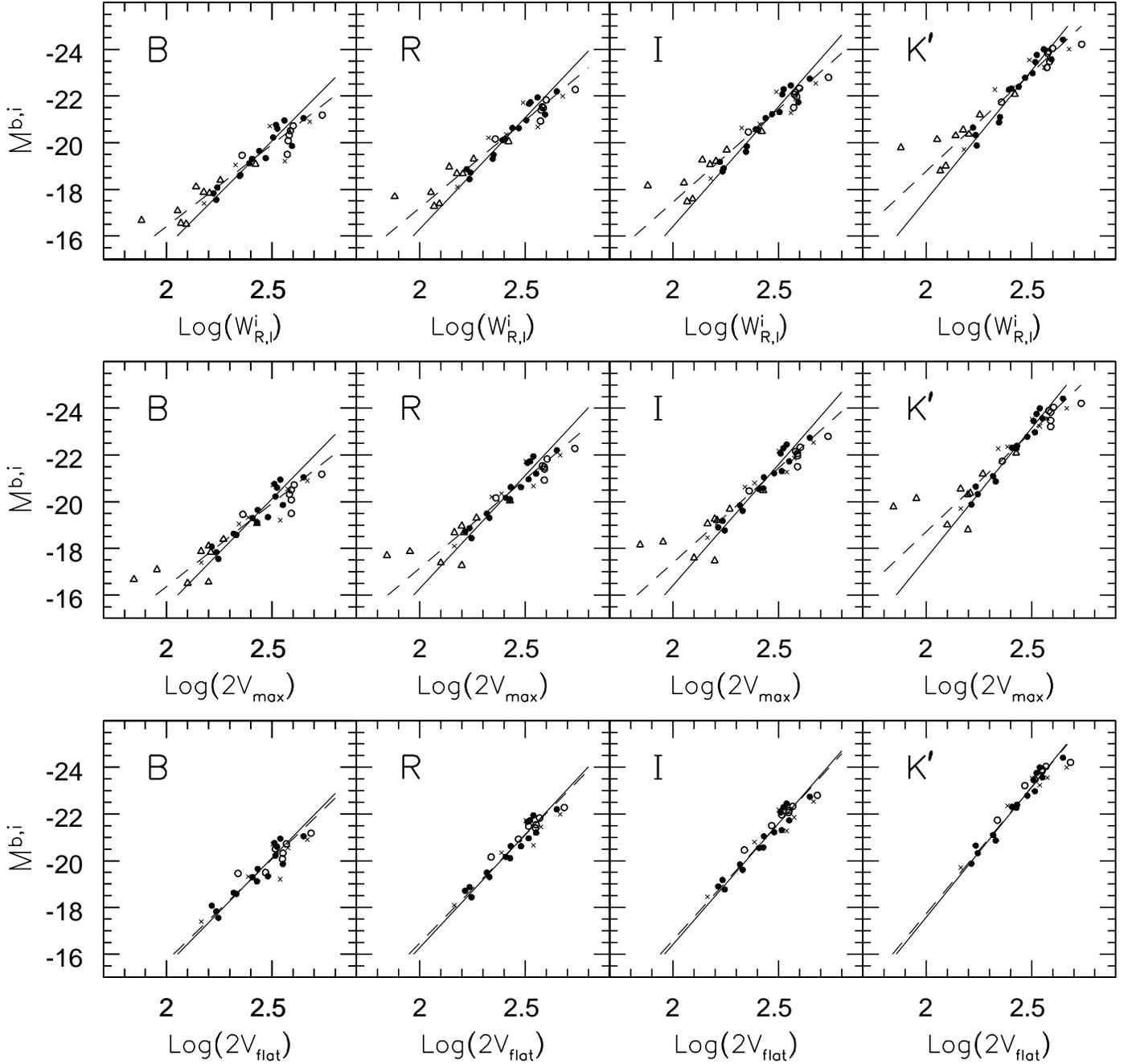}
\caption{TF-relations for all 31 galaxies in the \RCsample\ with
measured rotation curves (circles and triangles). The crosses indicate
the additional seven galaxies in the \SIsample\ for which a reliable
rotation curve could not be measured. Those seven galaxies were
ignored when making the inverse least-squares fits. The TF-relations
are constructed for each of the 4 available passbands using 3
different kinematic measures; the corrected width \WRIi\ of the global
HI profile (upper row), the maximum rotational velocity \Vmax\
measured from the HI rotation curve (middle row) and the amplitude of
the flat part \Vflat\ of the HI rotation curve (lower row). The open
triangles indicate galaxies with R-curves, the open circles indicate
galaxies with D-curves and the filled symbols indicate galaxies with
F-curves. Solid lines show the fits to the 15 filled circles only. The
dashed lines show fits using all galaxies from the \RCsample\ in each
panel.}
\end{figure*}

\begin{figure}
\epsscale{0.48}
\plotone{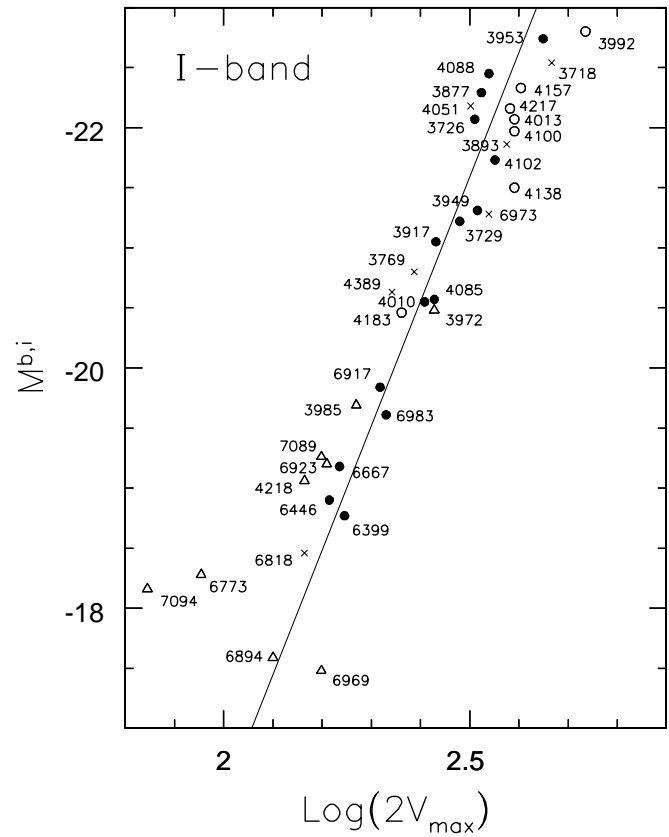}
\caption{Enlargement of the I-band panel in the middle row of
Figure~5.  Here, the symbols are labeled with a galaxy's NGC or UGC
number to allow checking with the photometric and HI synthesis data of
Papers I and IV.  NGC numbers run from 3718 through 4389, UGC numbers
run from 6399 through 7094.}
\end{figure}

\begin{figure*}
\epsscale{1.00}
\plotone{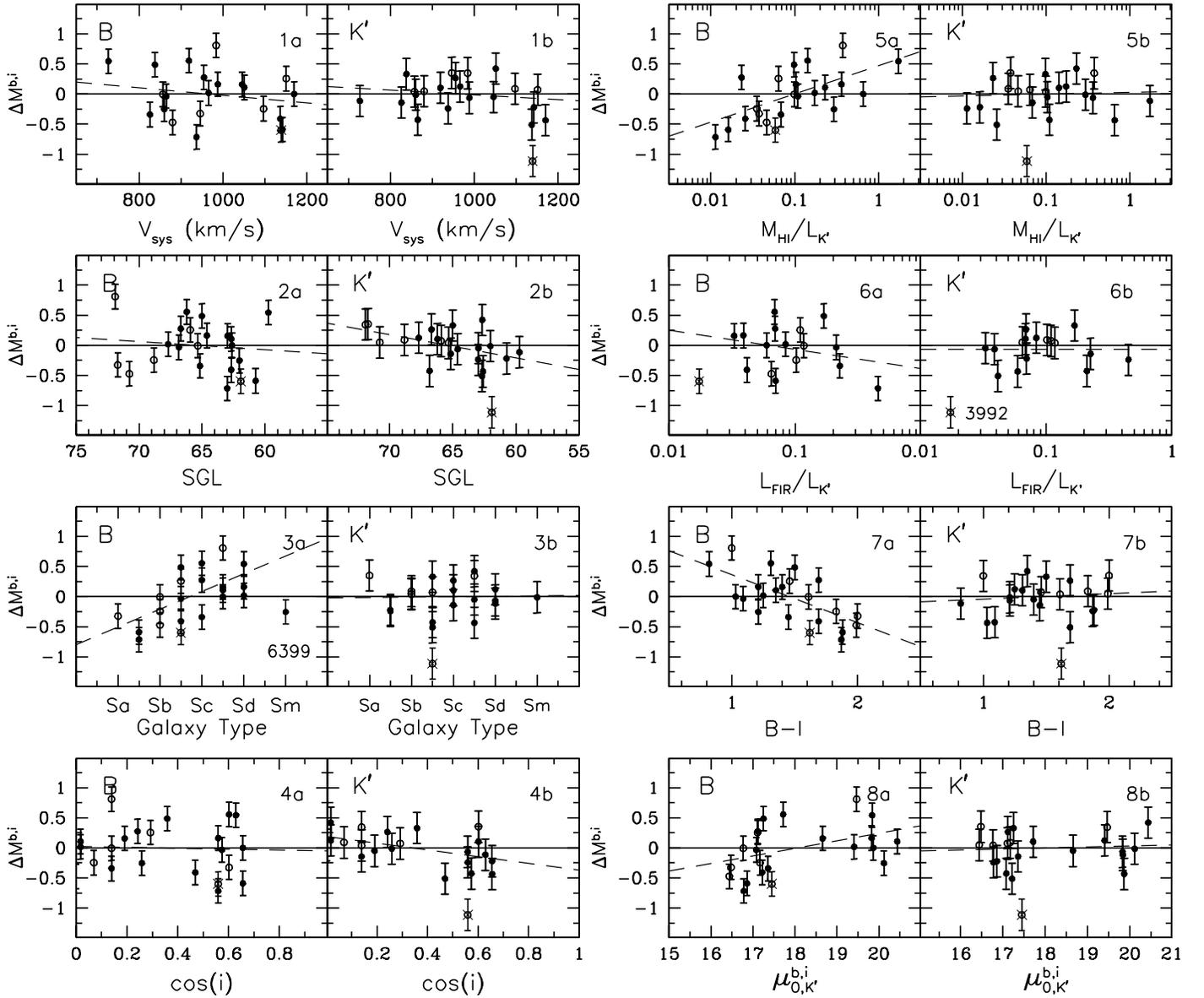}
\caption{Residuals of the M$^{\mbox{\scriptsize
b,i}}_{\mbox{\scriptsize B,\Kp}}$$-$Log(2\Vflat) TF-relations as a
function of various global properties of the spiral galaxies: 1)
systemic velocity, 2) Super-Galactic Longitude, 3) morphological type,
4) inclination angle, 5) relative HI content, 6) relative dust
content, 7) global B-I color and 8) face-on central \Kp\ surface
bightness of the fitted exponential disk. Only the 21 galaxies in the
\RCFDnoNGCsample\ are considered for the fits. The excluded galaxy NGC
3992 is indicated by the crossed out open circle. U6399 was excluded
from the fit in panel 3a.  Filled circles: galaxies with \Vmax\
=~\Vflat\, open circles: galaxies with \Vmax~$>$~\Vflat.}
\end{figure*}

\begin{figure*}
\epsscale{1.00}
\plotone{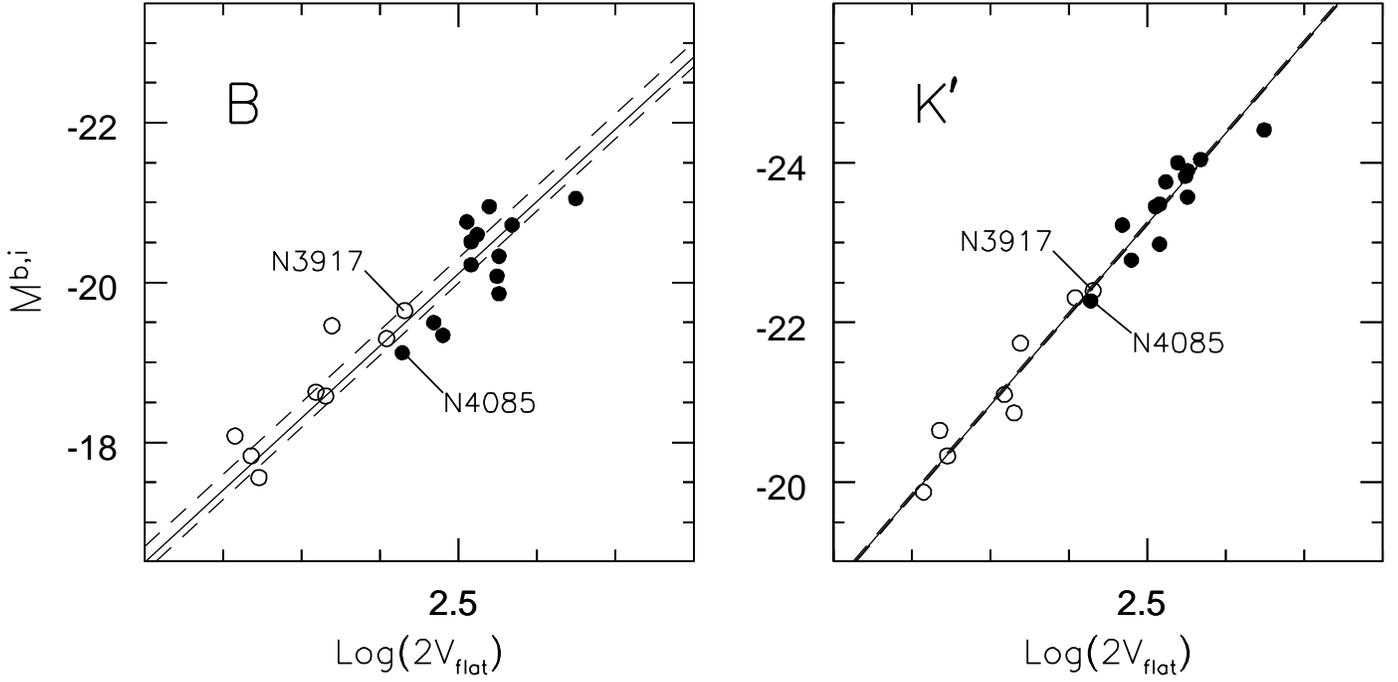}
\caption{The M$^{\mbox{\scriptsize b,i}}_{\mbox{\scriptsize
B,\Kp}}$$-$Log(2\Vflat) relations for the \RCFDnoNGCsample. In this
case, the filled circles indicate the HSB galaxies and the open
circles correspond to the LSB systems. The solid lines correspond to
inverse fits to all the galaxies. The dashed lines show the fits to
the HSB and LSB subsamples separately, keeping their slopes fixed to
those of the solid lines. The HSB/LSB galaxy pair N3917/N4085
discussed in the text is indicated}
\end{figure*}

\begin{figure}
\epsscale{0.48}
\plotone{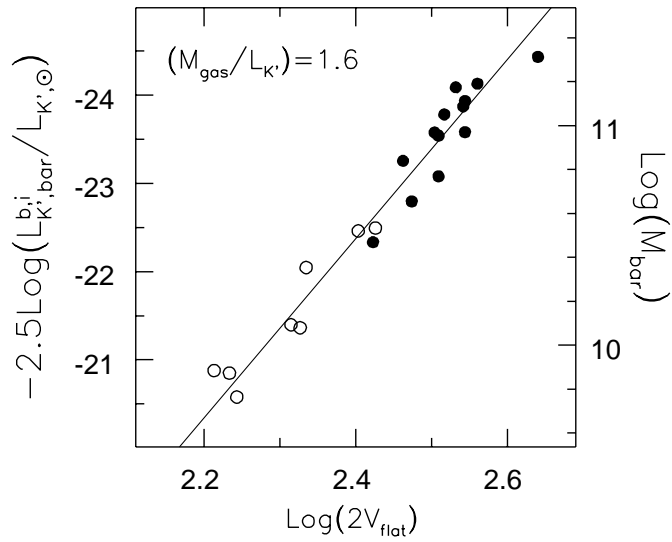}
\caption{The baryonic Log(${\cal M}$$_{\mbox{\scriptsize
bar}}$)$-$Log(2\Vflat) TF-relation for the \RCFDnoNGCsample\ assuming a
mass-to-light ratio of 1.6 for the gas. The solid symbols indicate HSB
galaxies, open symbols indicate LSB systems. The solid line shows an
inverse least-squares fit to all the points and has a slope of
$-$10.0.}
\end{figure}

\end{document}